\documentclass{elsart3p} 
\usepackage{natbib}
\bibpunct{[}{]}{;}{a}{,}{,} % customizing Harvard style referencing 
\usepackage{graphicx} 
\usepackage{amssymb} 
\usepackage{lineno} 
\begin{document} 
%-------------------------------------------------------------------------
 
%-------------------------------------------------------------------------
\begin{frontmatter} 
 
 
\title{Potential of Radiotelescopes for Atmospheric Line Observations:\\
       I. Observation Principles and Transmission Curves for Selected Sites} 
%\thanks[unklar]{} 
%\author{Nicola Schneider \corauthref{cor1}} 
\author[label1]{Nicola Schneider\corauthref{cor}},
\ead{nschneid@cea.fr} 
\author[label2]{Joachim Urban},
\ead{joaurb@chalmers.se} 
\author[label3]{Philippe Baron} 
\ead{baron@nict.go.jp} 

%\author{Nicola Schneider\thanksref{label1}} 
%\ead{nschneid@cea.fr} 
%\author{Joachim Urban\thanksref{label2}} 
%\ead{jo.urban@rss.chalmers.se} 

%\thanks[label2]{} 
%\corauth[cor1]{} 
%\address{Address\thanksref{label3}} 
%\thanks[label3]{} 
 
%\author[label1,label2]{} 
\address[label1]{IRFU/SAp CEA/DSM, Laboratoire AIM CNRS - Universit\'e Paris 
        Diderot, F-91191 Gif-sur-Yvette, France} 
\address[label2]{Chalmers University of Technology, 
              Department of Radio and Space Science, 
              412~96~G\"oteborg, Sweden} 
\address[label3]{National Institute of Information and Communication Technology, Koganei, 
                 Tokyo, Japan} 
\corauth[cor]{Corresponding author. Tel.: +33 169 08 2266; fax: +33 169 08 6577}

\begin{abstract}
  Existing and planned radiotelescopes working in the millimetre (mm)
  and sub-millimetre wavelengths range provide the possibility to be
  used for atmospheric line observations.  To scrutinize this
  potential, we outline the differences and similarities in technical
  equipment and observing techniques between ground-based aeronomy
  mm-wave radiometers and radiotelescopes.  Comprehensive tables
  summarizing the technical characteristics of existing and future
  (sub)-mm radiotelescopes are given. The advantages and disadvantages
  using radiotelescopes for atmospheric line observations are
  discussed.  In view of the importance of exploring the sub-mm and
  far-infrared wavelengths range for astronomical observations and
  atmospheric sciences, we present model calculations of the
  atmospheric transmission for selected telescope sites
  (DOME-C/Antarctica, ALMA/Chajnantor, JCMT and CSO on Mauna
  Kea/Hawaii, KOSMA/Swiss Alpes) for frequencies between 0 and
  2000\,GHz (150 $\mu$m) and typical atmospheric conditions using
  the forward model MOLIERE (version~5). For the DOME-C site, the
  transmission over a larger range of up to 10 THz (30 $\mu$m) is
  calculated in order to demonstrate the quality of an earth-bound
  site for mid-IR observations. All results are available on a dedicated webpage.
\end{abstract} 
 
 
\end{frontmatter} 
%-------------------------------------------------------------------------
 
% start of main text 

%-------------------------------------------------------------------------
\section{Introduction} 
\label{intro} 
 
\subsection{Atmospheric line observations with radiotelescopes} 

%General: submm/mm wavelengths range 
The millimetre and sub-millimetre wavelengths range contains a large
number of atmospheric emission lines of molecular transitions 
(rotational, rotation-vibration, hyperfine structure lines), 
including important atmospheric species such as H$_2$O, O$_3$,
O$_2$, N$_2$O, HCl, ClO, OH, CO, among many others.  
Atmospheric scientists observe these molecules  
in different geometries (limb-, nadir-, and 
zenith-sounding) either with ground-based
mm-wave radiometers in the frequency range of 20 to 300\,GHz
(15 to 1\,mm) or with airborne and spaceborne sensors (up to
$\sim$2.5\,THz or 110\,$\mu$m), less affected by the absorption of
water vapor in the Earth's troposphere. While the references for
observations in the mm-wavelengths range are numerous, there are
significantly fewer observation systems at high frequencies.
%since this usually requires airborne and spaceborne instrumentation.  
Examples for satellite instruments are SMR (Sub-Millimetre Radiometer)
on Odin \citep{frisk2003}, launched in 2001, working in the frequency
band 486--580 GHz and the Microwave Limb Sounder (MLS) on Aura,
operating several radiometers around 180, 230, 625--660 GHz, and at 2.5
THz since 2004 \citep{waters2006}. The launch for JEM/SMILES (Japanese
superconducting submillimeter-wave limb-emission sounder) with a 640
GHz SIS radiometer is planned for late 2009 \citep{kasai2006}.

% balloon-borne SMILES ?

% From ODIN website
% (http://www.aerospace-technology.com/projects/odin/) 
%SPECTRAL LINE
%The radiometer package covers transitions of aeronomical interest from
%the following molecules: ClO, CO, NO2, N2O, H2O2, HO2, H2O, H218O, NO,
%N2O, HNO3, O3 and O2 as well as atomic and molecular transitions of
%astrophysical interest from: CI, H218O, H2O, H2S, NH3, H2CO, O2, CS,
%13CO, H2CS, SO, SO2. The optical spectrometer is aimed at studying the
%following species in the Earth atmosphere: Aerosols, ClO, O3, O2, O4,
%NO and NO2
%The OH was observed at 1.8 and 2.6 THz from aircraft by xxx. Trace gas
%concentrations of higher excited lines of OH, HO$_2$, ClO, BrO and
%other key molecules of the Earth's stratosphere will be observed at
%500, 650, and 1800 GHz onboard TELIS (``TErahertz LImb Sounder''), a
%balloon-borne experiment (reference). {\bf Nicola: Are there other
%examples of radiometer observations at frequencies higher than 500 GHz
%?}

% Referenz fuer den THZ teil: \citet{mair2007}.
% is currently under development at the German Aerospace Center (DLR) 

%Radiotelescopes fig.n aeronomy observations performed with them 
Ground-based microwave radiotelescopes (RTs from now on) are normally
devoted to perform radio-astronomical observations in the (sub)-mm
region which is the typical emitting wavelength range for 
various transitions of molecules. More than 140 molecules have
already been detected in space (with CO as the most abundant one after
H$_2$) and the list is continuously growing.  Planned facilities at
extremely dry sites, such as DOME-C in Antarctica or CCAT
(Cornell-Caltech Atacama Telescope) in the Andes, would enable to
access even the THz range and thus observations of interesting atomic
and ionic lines like ionized nitrogen, [NII], at 1.46 THz, 
and molecular transitions like high-J CO,
HCN, and HCO$^+$ lines and light hydrides (NH$^+$, NH$_2$ etc.)  would
become possible. These tracers are used to study the physics of UV
illuminated regions or very dense star forming regions. 

% atomic an ionic lines: [NII], [CII], [CI] 
% [NII] 3P2 - 3P1   2.46 THz, 122 mu
%       3P1 - 3P0   1.46027 THz, 205 mu
% [NII] probes low-density ionized gas, extinction free, complement to 
% [OIII] 88 and 52 mu tracing high density ionized gas 
% [CII] at 1901.28, 158 mu
%
% high-J CO, HCN, HCO+ rotational transistions 
%
%  HCN 9-8  797 GHz
%      10-9 886 GHz
%
%  HCO+ 9-8 802 GHz 
%      10-9 892 GHz  
%
% also H13CN, HC15N, HCN-v2, DCN
%
% higher frequencies (for SOFIA calculations) : 
% HD 2674.99 GHz
% OI 4744.66 GHz 
%
% Light Hydrides: H2D+, D2H+, NH2, NH+, H2O isotopes, LiH3
% p-H2D+ 1.37 THz
% NH+  1.0126 THz 
% NH2  462 GHz, 902 GHz 
%% low bending modes of carbon chains (200 and 230 mu window)

% There is a 'HIFI' gap between 1250 and 1410 GHz 

%RT alrady used for atmospheric line observations
The technical equipment and observing procedures are originally
designed for astronomical observations that normally cancel out the
atmospheric contribution. In addition, RTs always work in up-looking
geometry.  However, using special observing methods, RTs already
have been used in the past to make sporadic measurements of the line
emission of atmospheric species (e.g.  H$_2$O, \citet{barret1962,
groom1965, bevilaqua1983}; HO$_2$, \citet{sandor1998}; O$_3$,
\citet{caton1968}, \citet{connor1987}; ClO, \citet{parrish1981},
\citet{solomon1984}; CO, \citet{waters1976}), and to perform long-term
monitoring projects (e.g.  CO, \citet{boes1994}). O$_2$-Zeeman
splitting was observed by \citet{pardo1995}.  The typical altitude
range covered by these observations is 10--80\,km with an altitude
resolution ranging from 5--20\,km. Going down much lower in altitude 
is not always useful because (i) the molecular line is very broad and 
spectrally not resolved due to the limited bandwidth of the receiver, 
and (ii) absorption (= emission) is dominated by water vapor and dry air and 
thus a line can not be separated from the water vapor and dry air continuum 

%Why to observe atmospheric tracers ? 
The motivations for observing atmospheric lines are mannifold. Many
mm-wave heterodyne instruments belong to the ``Network for Atmospheric
Composition Change'' (NDACC, http://www.ndsc.ncep.noaa.gov). These and
other radiometers focus on the short and long term abundance
variations of stratospheric ozone \citep[e.g.][]{sinnhuber1998,
calisesi2001, klein2002, schneider2003, raffalski2005,
steinbrecht2006} as well as strato-mesospheric water vapor
\citep[e.g.][]{seele1999, nedoluha2003}. A large number of molecules
of prime interest for middle atmospheric sciences, often having only
very low abundances, can be observed and studied using this remote
technique. Besides, (mainly campaign-based) observations of key
species related to polar ozone depletion such as ClO and HNO$_3$
\citep{shindell1996, dezafra2001, muscari2002}, and longer-lived
species such as N$_2$O and CO have been succesfully measured. They are
particularly useful as tracers of middle atmospheric transport
\citep[e.g.][]{crewell1996, dezafra2004, forkman2005}.  Another
example is the observation of the atmospheric HCN molecule that has
been detected at three different radio-astronomical observatories,
i.e.  the {\em Cologne Observatory for Sub-mm Astronomy} (KOSMA,
Gornergrat, Switzerland), the {\em James Clark Maxwell Telescope}
(JCMT, Mauna~Kea, Hawaii), and the {\em Caltech Submm Observatory}
(CSO, Mauna~Kea, Hawaii) \citep[e.g.][]{despois2000, lautie2003}.
These measurements will be further discussed in another paper
[Lauti\'e et al.  2009 in prep., Paper-II].  A dedicated instrument
routinely used on an astronomical telescope (the CSO) for atmospheric
studies is a Fourier Transform Spectrometer \citep{pardo2004}.

RTs are also powerful tools to study the chemistry in the upper
stratosphere and mesosphere \citep{sandor1998}. They can record the
diurnal variations of atmospheric trace gases and, hence, complete
observations from sun-synchronous orbiters, which are at fixed local
times. Furthermore, observations with RTs can provide a database of
rarely observed molecules to validate satellite observations  
and can be used for observations of the atmosphere of other planets 
in the solar system.
Finally, RTs can also be used to derive mesospheric wind vectors from
the Doppler-shift of observed lines \citep{burrows2007}.

%Due to the signal absorption of tropospheric water vapor, which
%increases with frequency throughout the mm- and sub-mm wavelengths
%range, sufficiently low optical thicknesses for ground-based
%observations of middle atmospheric target species can mainly be found
%at frequencies smaller than $\sim$300\,GHz (wavelengths larger than
%$\sim$1\,mm).

Due to the increase of the tropospheric background absorption
(dominated by water vapor and dry air continuum) with frequency
throughout the mm- and sub-mm wavelengths range, spectral windows with
sufficiently low optical thicknesses for ground-based observations of
middle atmospheric target species can mainly be found at frequencies
smaller than $\sim$300\,GHz.  
Since, at the same time, the line intensities of most
minor species increase with frequency in the sub-mm range, highest
possible frequency windows are of interest and measurements shall
preferably be performed at extremely dry sites, similar to
radio-astronomical observations. 
However, the maximum upper limit for obtaining a profile and not only
a column, i.e. the altitude of the transition from pressure to Doppler
broadened lines, decreases with rising frequency.  It is typically 90
km at 100 GHz and 70 km at 2 THz.

%However, the maximum upper limit for obtaining a profile and not only
%a column decreases with rising frequency. It is typically 90 km at 100
%GHz and 70 km at 2 THz.

%What we do in this paper
In order to efficiently use RTs for atmospheric line observations, it
is mandatory to have an overview of available front- and back-ends of
existing and planned RTs and to fully understand the employed
observing techniques.  For this reason we present in this paper a
short review of the basic principles how radio-observations are
performed and summarize in detail the main properties of the
instruments.
%(frequency-ranges, bandwidths, spectral resolution etc.).
Comprehensive tables including all astronomical (sub)-mm facilities
are given.
%This review includes observatories and interferometers in construction
%(e.g.  ALMA/Chajnantor,  LMT/Mexico) and new  sites that are currently
%being tested (e.g. DOME-C in Antarctica).

\subsection{Transmission curves} 

%Transmission curves for different sites
For both, astronomical and aeronomy observations in the
sub-mm/far-infrared (FIR) range, it is of prime importance to assess
the quality of the atmosphere in terms of transparency.  Only
extremely dry sites allow to access this spectral range.  We therefore
used the forward model MOLIERE-5 \citep{urban2004a} to predict the
atmospheric transmission for sites of interest for sub-mm/FIR
astronomy.  In this paper, we show the results for frequencies between
0 and 2000\,GHz for the sites of Chajnantor/Chile, DOME-C/Antarctica,
Mauna Kea/Hawaii, and KOSMA/Swiss Alps.  A more comprehensive list of
observatories covering different wavelengths and various regimes of
atmospheric conditions is made available on
http://transmissioncurves.free.fr.  In Paper-III of this series
(Schneider et al., 2009, in preparation) we compare the results
obtained with MOLIERE-5 to other atmospheric models used in
radioastronomy, i.e. ATM (Atmospheric Transmission at Microwaves,
\citet{pardo2001}), ATRAN (Atmospheric TRANsmission,
\citet{lord1992}), and AM (Atmospheric Model, \citet{paine2004}).

%-------------------------------------------------------------------------
\section{Radiotelescopes vs. mm-wave radiometers} 

In this section, we summarize the principal similarities and
differences between mm-wave radiotelescopes and aeronomy radiometers.
An overview of the sites is given in Table~\ref{table1} (as of summer 
2009 see individual websites given in column 6 for more recent
information), and the main instrumental characteristics are summarized
in Table~\ref{table2}.

% {\bf  Nicola:  Wir koennten auch  eine  Tabelle mit fest stationierten
%  mm-Empfaengern fuer Aeronomy machen ? Was meinst Du ? Welche gibt es
%  da ueberhaupt ?}\\
%
% {\bf Jo: possible - let's see later whether this fits. Information can
%  be   found   at      the    NDACC   microwave       group  home-page
%  \begin{verbatim}
%  http://www.iapmw.unibe.ch/research/collaboration/ndsc-microwave/instruments/
%  \end{verbatim}
%  One could simply indicate the web-page as reference ... 
%  anyway, a user will not be able to use the instruments (just the
%  data which are available from the NDACC web-site!}\\

\subsection{Technical equipment}
\subsubsection{Optics} 
Millimeter-wave RTs are movable, single, parabolic dishes with 1 to
currently 45\,m diameter.  The emission is received in the
main dish and reflected on the subreflector.  The {\em Half Power
Beam Width} (HPBW) depends on the telescope diameter d and the
observing frequency $\nu$ (HPBW\,[$'$] $\sim$ 1260/(d [m] $\times$ \,
$\nu$ [GHz]) and varies typically between 0.1$'$ (300\,GHz\,@\,30m)
and 11$'$ (100\,GHz\,@\,1\,m). \\ In contrast, aeronomy heterodyne
receivers are normally non-steerable instruments with a focusing
mirror or horn antenna as entrance aperture, leading to large beam
sizes of typically a few degrees.  The size of the beam as well as a
possible error-beam pickup is normally unimportant for atmospheric
observations.

\subsubsection{Front-end}  \label{frontend} 
Both types of instruments use the superheterodyne principle
\citep{kraus1986} to convert the high frequency signal f$_S$ via a
mixer into a low-frequency intermediate signal f$_{IF}$.  Due to the
mixing process with a signal from a local oscillator f$_{LO}$, two
major frequencies are produced: f$_{LO}$--f$_S$ and
f$_{LO}$+f$_S$.  Heterodyne receivers can work in double- (DSB) or
single-sideband mode (SSB), depending on whether the detector
receives both side bands or whether one is suppressed.  In the latter
case, one of the bands f$_{LO}$--f$_S$ or f$_S$--f$_{LO}$ can be
choosen.  Aeronomy line observations are normally performed in SSB
mode in order to prevent confusion of pressure-broadenend lines from
both sidebands.

Critical parameters for the detection of a signal T$_{sig}$ [K] are
the frequency resolution $\Delta\nu$ [Hz], system noise temperature
T$_{sys}$ [K] and integration time $t_{int}$ [s]. We define the system
noise temperature as the sum of the receiver noise temperature and the
signal brightness temperature: T$_{sys}$ = T$_{rec}$ + T$_{sig}$ [K].
The rms noise is $\Delta$T$_{rms}$\,=\,T$_{sys}$/$\sqrt{(\Delta\nu \,
t_{int})}$.  However, the radiometer formula is only correct as long
as the receiver noise is thermal in nature. The time when the
Allan variance \citep{allan1972} has a minimum ("Allan minimum time"),
is the maximum integration time up to which the noise decreases with
integration time. The integration time should therefore not be longer
than this Allan minimum time. It is typically in the range of a few
seconds up to several tens of seconds.

The receiver temperature depends critically on the mixer and 
first amplifier in the chain.  {\em Schottky-diode} mixers use the
non-linear current-voltage characteristic of metal/semi-conductor
diodes for the mixing process and reach typically noise temperatures
of a few hundreds to one thousand Kelvin (double sideband mode, 100 to
400\,GHz).  SIS ({\em Semiconductor-Insulator-Semiconductor})
receivers on the other hand, reach DSB noise temperatures  
as low as around 80 to 200\,K in the same frequency range.

RTs nowadays employ SIS receivers up to 1 Terahertz (THz) which
are cooled by a reservoir of liquid helium or a closed-cycle system.
The same is true for stationary aeronomy radiometers though Schottky
systems are still employed due to their easy handling and
robustness. This is an important criteria for automatic operation over
long time periods in remote locations. Many of these receivers are
tuned to a fixed frequency in order to monitor a particular spectral
line (e.g O$_3$, H$_2$O).  Recent developments to reach higher
frequencies --\,both, in aeronomy and astronomy\,-- employ {\em Hot
Electron Bolometers} (HEB) that operate as heterodyne mixer elements
for down conversion of the observed signal into an intermediate
frequency signal. For example, the HEB instrument CONDOR (CO N$^+$
Deuterium Observations Receiver,
\citet{wiedner2006},\citet{wieching2007}) working between 1.3 and
1.5\,THz, was successfully installed at the APEX radiotelescope in
Chile.

\subsubsection{Back-end} 
Spectroscopy of (sub)mm-wave molecular and atomic lines requires a
real-time spectrometer in combination with the heterodyne receiver.
Different types of such spectrometers are available, namely
'filterbanks', 'chirp-transform spectrometers (CTS)',
'auto-correlators' (AC), 'acousto-optical spectrometers (AOS)', and
'FFTS' (Fast Fourier Transform Spectrometer).
%, e.g. \citep{klein2006}).
While CTS 
%\citep{hartogh1990} 
offer only relatively small bandwidths, all other instruments have
been built with bandwidths of the order of 1--4\,GHz which is
important for the detection of pressure-broadened atmospheric lines in
order to cover the contributions to the spectrum originating from low
altitudes (typically a few to 20 km).  An equidistant resolution
over the whole bandwidth is usually required.  Upper atmosphere
sounding requires a frequency resolution of the order of 0.1--1
MHz. Bandwidth and spectral resolution of a filterbank spectrometer
are in principle flexible design parameters, however with increasing
bandwidth and high spectral resolution this type of instrument is
getting voluminous and unpractical.  Thus, AOS, AC, and FFTS are often
the preferred choice for both, aeronomy and astronomy, measurement
systems.

\subsection{Observing Methods}

\subsubsection{Radiotelescopes} \label{observing-rt}
\begin{figure*}
  \begin{center}
  \includegraphics[width=7cm,angle=0]{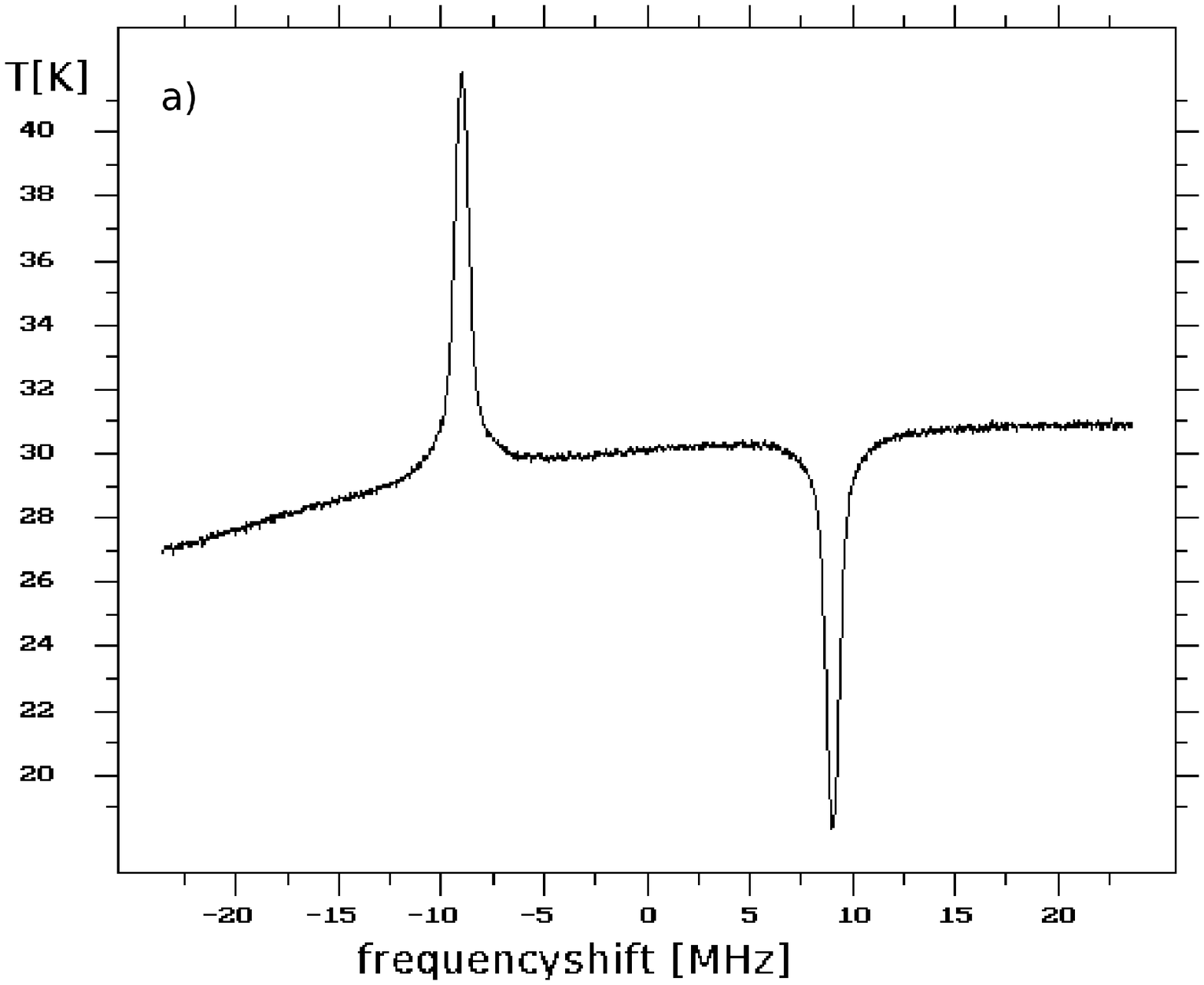} 
  \includegraphics[width=8cm,angle=0]{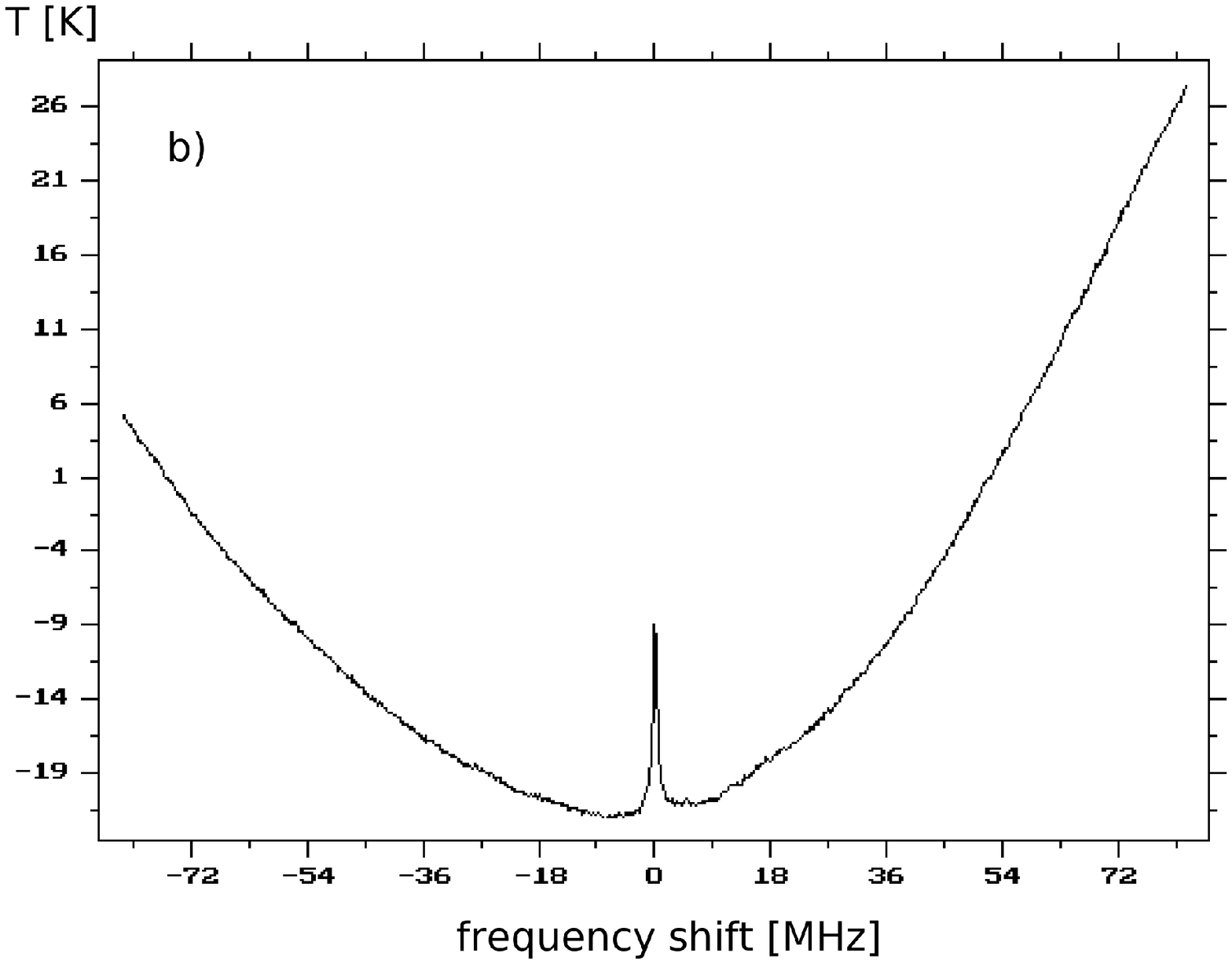}
  \includegraphics[width=7cm,angle=0]{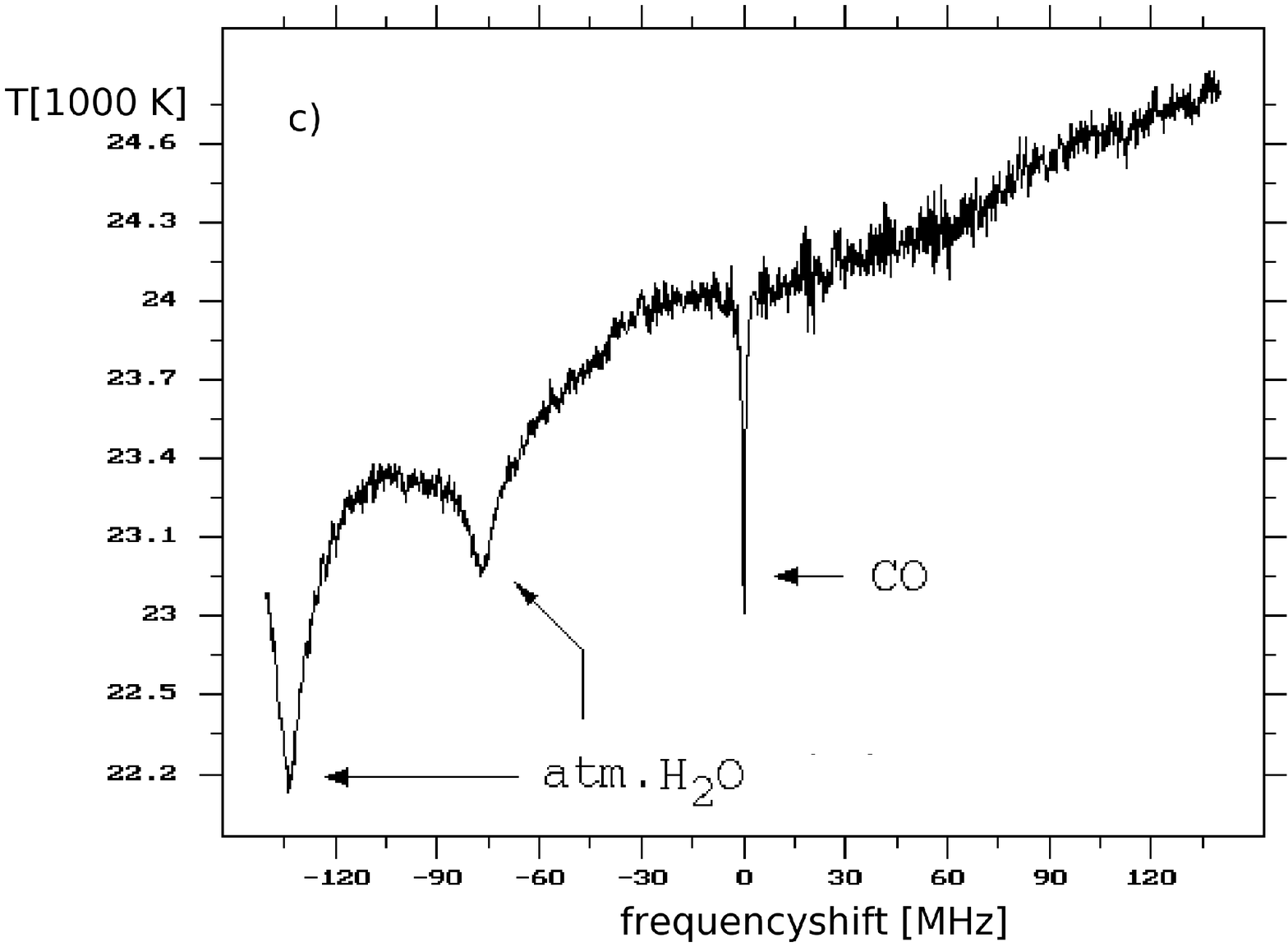}
  \includegraphics[width=7cm,angle=0]{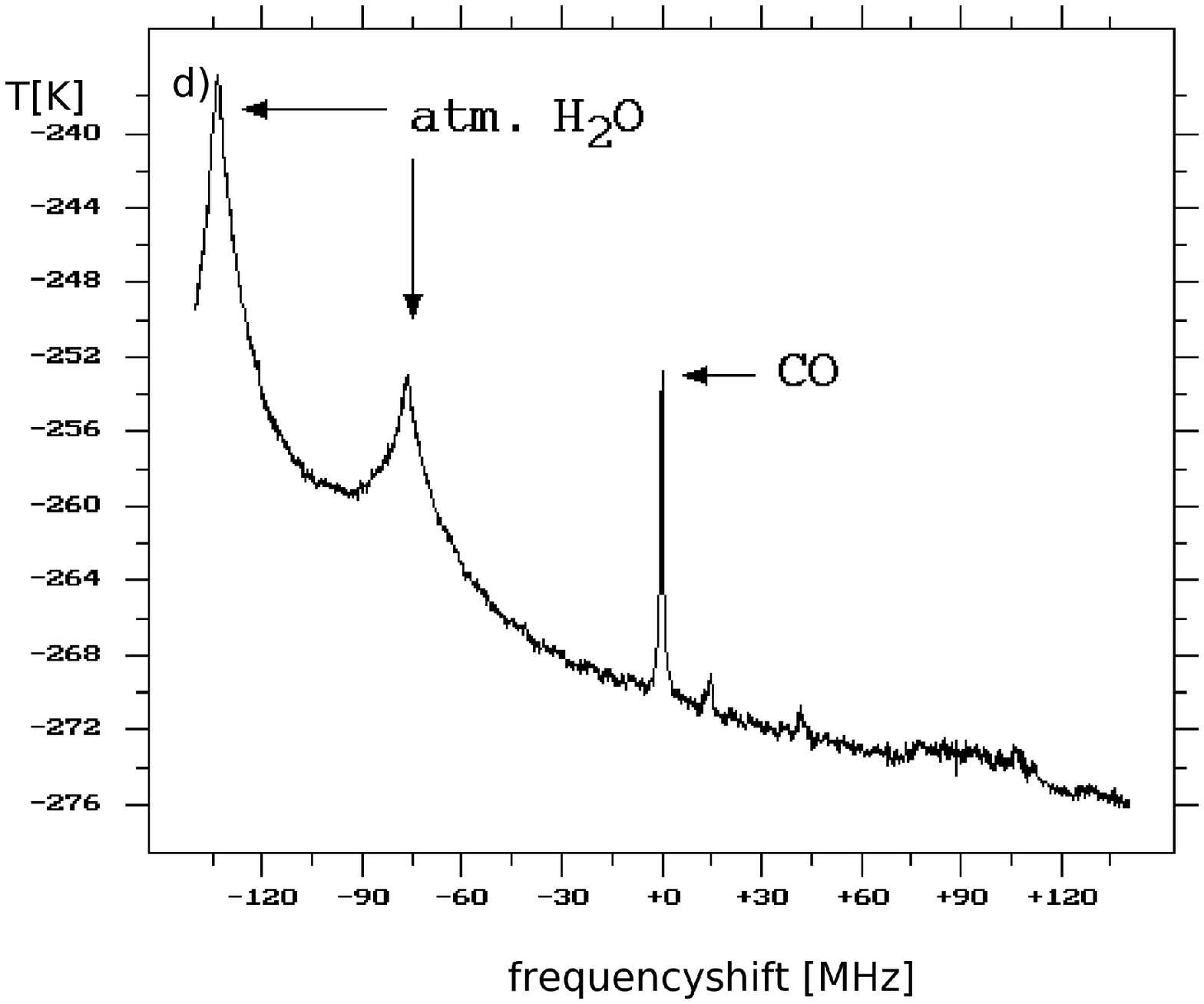}
  \end{center}
  \caption{Examples of mesospheric $^{12}$CO 2$\to$1 measurements at 
   230 GHz, taken with the KOSMA 3m telescope \citet{boes1994}. All observations 
   were performed with a Schottky receiver (T$_{rec} \sim$400 K) and an acousto 
   optical spectrometer (AOS) with a bandwidth of 287 MHz and a spectral 
   resolution of 167 kHz. The integration time is typically one hour.  
   a) Simple frequency switching observation, b) load swichting 
   observation using the cold load as reference, c) absorption measurement against the sun, and 
   d) absorption measurement against the mountains. }
   \label{fs}
\end{figure*}

The definition for observing with RTs is that a {\sl signal} position
refers to the emitting astronomical source and a {\sl reference}
position to an emission free position away from the astronomical
source. In the following, we give an overview of all observing methods
employed for observations using RTs. Figure~\ref{fs} shows
measurements of mesospheric CO, taken at the KOSMA 3m telescope, to
illustrate the observing methods suitable for atmospheric line
measurements.

\paragraph*{Position-switching (PS)\\} 
This is the most simple observing mode for RTs.  The telescope
observes alternately the signal and the reference position, the latter
being subtracted from the signal. This procedure eliminates most
instrumental and atmospheric effects, accordingly it is not suitable
for telluric lines.  The integration time may be adapted to the total
system stability time determined by the Allan variance method
\citep{allan1972}.  Standing waves in the optical path appear as
frequency dependent baseline ripples in the spectrum.  Moving the
sub-reflector between the signal- and reference phases by 1/4 of the
observing wavelength can help to compensate for this effect.  We
succesfully employed this correction-method for our frequency-switching 
atmospheric HCN observations at the KOSMA radiotelescope (Paper II).

%However, ripples produced elsewhere in the system,  for example due to
%cable flexures, can not be improved by focus modulations.

\paragraph*{Beam Switching (BS)\\}  
For this observing mode, the secondary mirror is tipped by a small
angle ('chopped') at a choosen frequency and amplitude ('throw'). The
detected signal is the difference between the sky and reference
signals. The advantage of this method is that switching can be
performed much faster than with the PS method but the switch throw is
normally limited in distance to the source (up to a few
10$'$). This method also naturally cancels out atmospheric lines.

\paragraph*{Frequency Switching (FS)\\}  
In this case, the frequency of the local oscillator is switched
between two values of the LO frequency (f and f+$\delta$f), separated
by typically a few to a few tens of megahertz, whilst observing only
in the signal path.  Both spectra (S(f) and S(f+$\delta$f)) are then
subtracted so that in the final spectrum
S1\,=\,S(f+$\delta$f)\,--\,S(f), a positive and a negative line appear
(see Fig.~\ref{fs}a).  A modification of simple FS is the 'Double
frequency switching' mode in which the LO is also switched in the
opposite direction (f--$\delta$f), leading to an additional spectrum
S2\,=\,S(f)\,--\,S(f--$\delta$f).  The final spectrum can then be
computed as S\,=0.5\,(S1\,--\,S2)
\,=\,S(f)\,--\,0.5$\times$S(f--$\delta$f)\,--\,0.5$\times$S(f+$\delta$f).
 
If the LO system (normally stabilized by a phase locked loop) is not
stable enough to support fast switching, FS can be mimiced by
switching the frequency every signal phase (typically 20--40\,s).  All
FS procedures enable to observe atmospheric lines directly.  However,
if a frequency range with a high density of spectral lines is
observed, crowding and overlapping of spectral features can be a
problem, in particular if working in DSB mode.  Broad lines such
as those from ozone in the lower stratosphere are very critical whereas
narrow lines like mesospheric CO are less critical. If the frequency
throw is too small, the line wings of a pressure-broadened line are
cut off, restricting the exploitable altitude range.

\paragraph*{Load Switching (LS)\\}  
If the reference position is not taken on the sky, as for total
power, but on the cold load of the receiver, formally a spectrum can
be calculated by subtracting the load spectrum from the signal
spectrum (see Fig.~\ref{fs}b).  Atmospheric lines are generally
observable with this method but the continuum offset between cold load
(liquid nitrogen at 77\,K) and sky is large and standing waves are
typical so that this method is not appropriate for detecting weak
lines.

\paragraph*{Absorption measurements (AM)\\}  
Absorption measurements against Sun, Moon or mountains are performed
in a similar way as for load switching by replacing the cold load with
the astrophysical object or mountain (see Fig.~\ref{fs}c and
~\ref{fs}d). However, the calibration of these spectra is
often difficult due to large differences in the continuum level and
due to the fact that neither mountains nor planets, moon, or sun are
perfectly emitting as a blackbody. 

\subsubsection{Aeronomy Receivers} 
\vskip0.5mm

\paragraph*{Total Power \\} 
Strong atmospheric lines can be observed using a total power method,
i.e.  the sky measurement is performed at a fixed elevation angle and
the received signal is calibrated in frequent intervals with
measurements of a hot and cold calibration source.  As explained
above, baseline ripples due to standing waves in the optical paths may
be suppressed by using a phase-wobbler which frequently changes the
optical path lengths (e.g. by 1/4 of the observing wavelength).

\paragraph*{Load switching and Frequency Switching \\}
These procedures (described in Sec.~\ref{observing-rt}) may also be
used for aeronomy observations.  See for example \citet{ricaud1991}
for a detailed description and application of the frequency-switching
technique for ozone observations as well as \citet{lobsinger1984} for
load switching ozone measurements.

\paragraph*{Balanced Beam Switching (BBS)\\} 
This commonly employed differential observing method was initially
developed by \citet{parrish1988}.  It uses an atmospheric
measurement at low elevation (typically between 10$^{\circ}$ and
20$^{\circ}$) resulting in a large emission path length along the
line-of-sight, providing a strong signal, and a reference measurement
(R) at higher elevation (typically $>$50$^{\circ}$) with a much weaker
emission strengths. In order to compensate for the offset in the
received power between the two views, an absorber (such as a
plexiglass plate) is put in the reference path.  Fine balancing of the
power levels in signal and reference path is achieved by varying the
elevation angle of the signal view. The difference S--R then contains
the differential signal emission in which --\,to first order\,--
instrumental effects such as receiver non-linearities and standing
waves are removed.  This technique is commonly employed by
ground-based radiometers within the framework of the NDACC.

\paragraph*{Balanced Load Switching (BLS) \\} 
An alternative to the beam switching technique is the {\em Balanced
Load Switching (BLS)} method, where the reference watches the cold
load instead of the atmosphere and the balancing is achieved by mixing
the cold signal with an absorber at ambient temperature using a
turnable wire grid \citep[e.g.][]{raffalski2005}. This 
procedure is similar to the 'Dicke-switching' method used in 
radioastronomy.

\subsection{Correction of tropospheric absorption}
\label{tropospheric_correction} 

Outside the line centres, the atmospheric transmission relevant to
an observer at the ground is determined by the tropospheric absorption
whilst strato-mesospheric lines and continua have typically only a
negligible contribution. The most important factor for the tropospheric
transmission is the (frequency dependent) tropospheric water vapor
profile. Atmospheric emission (and absorption) is a function of
temperature. Spectral line broadening as well as collision induced
non-resonant continuum absorption are also functions of pressure.

By using a mechanical or software simulated 'chopper-wheel'
calibration or a 'sky dip' measurement, the troposheric opacity $\tau$
is calculated and the detected emission in terms of calibrated antenna
temperature $T_A'$ [K] can be corrected as
\begin{equation}
 \label{tau}
T_A  =  T_A' \exp(-\tau \, A) \,\, [K]. 
\end{equation}
$A\approx 1/\sin(elevation)$ is the geometrical air mass factor.  The
'sky dip' method consists of measuring either distinct points or
continuously the atmospheric emission from high to low elevations and
derive the opacity by an exponential fit to equation \ref{tau}.  The
'chopper wheel' calibration \citep{penzias1973} defines the
calibration signal to be the difference between an absorber at ambient
temperature and the sky.  Finally, the calibrated antenna temperature
has to be corrected for the telescopes forward- and main beam
efficiency to derive main beam brightness temperatures \citep[see
e.g.][]{kraus1986}.

\subsection{Requirements for astronomical and atmospheric observations}
\label{needs} 

Astronomical observations with single dish RTs typically require
an absolute calibration accuracy of lower than 10\%, an equidistant,
good frequency resolution, and a good pointing accuracy (better than
20\% of the beamsize). The required beamsize depends on the
astronomical project, i.e.  large beamsizes for efficiently mapping
large areas on the sky, small beamsizes for extragalactic and detailed
Galactic studies.  For atmospheric line observations, the
calibration should ideally be better than 1\%, including sideband
ratio calibration and tropospheric correction. Pointing and beamsize
are less important but the requirements for long-term stability and
good baselines are high. Both, high-resolution frequency observations
(e.g. narrow lines in the mesosphere) and broadband observations
(broad lines in the troposphere and lower stratosphere) are important.

\subsection{Characteristics of Radiotelescopes}
Table \ref{table1} lists all radiotelescopes operating in the mm- and
sub-mm wavelengths range with an antenna radius between $\sim$1\,m and
45\,m which can potentially perform atmospheric observations.  Four of
them are located in Europe, five in the US (two of those in Hawaii),
three in South America (Chile), three in Asia, and one in
Australia. Eight telescopes are situated at high
altitude sites ($>$3000\,m).  Not listed are mm- and sub-mm
interferometers though it can be of interest to
use these explored sites to install aeronomy radiometers.  For
example, a radiometer monitoring ClO
\citep{ricaud2004} was placed on the Plateau de Bure in the French
Alps, hosting an interferometer with 6\,mm-antennas of 15\,m size
(http://iram.fr).  Other locations are the {\em Submillimeter Array}
(http://sma-www.harvard.edu/) in Hawaii/USA, the {\em Combined Array
for Research in Mm-wave Astronomy} (CARMA, http://www.mmarray.org)
located in California/USA, the {\em Australia Telescope Compact Array}
(ATCA, http://narrabri.atnf.csiro.au), and the Nobeyama Millimetre
Array (NMA, http://www.nro.nao.ac.jp).  The Atacama Large Millimetre
Array (ALMA) with at least 50 dishes of 12m diameter each is currently 
being constructed on the plateau of Chanjantor and is expected to 
be operational with some
antennas starting from 2012.  A 50m single dish sub-mm-RT (the Large
Millimetre Telescope, LMT, http://www.lmtgtm.org) is set up in Mexico
and foreseen to receive first light in 2009. Other sites under
investigation for sub-mm RT are located in Chile (the Cornell-Caltech
Atacama Telescope CCAT, http://www.submm.org) and Antarctica (DOME-C
at the Concordia Station, DOME-A at 80$^\circ$S on the Antarctican
Plateau at an altitude of 4093m, and DOME-F at 77$^\circ$S at an
altitude of 3810m).

The technical specifications of the telescopes listed in
Table~\ref{table1} are given in Table~\ref{table2}.  We do not
include bolometers in this list since we focus on
spectroscopy. However, using bolometers can be interesting for the study
of atmospheric emissivity changes due to pwv or broad planetary
absorption lines. For most of the telescopes,
different receivers in various tuning ranges are available.  
%For each radiometer the covered frequency range
%(column~2), the sideband operation mode (single or double) and mixer
%type (SIS or Schottky) (column~3), the typical noise receiver
%temperature (column~4) and receiver bandwidth (column~5) are given.
%The Intermediate Frequency is listed in column~6.  Column~7 to 9
%provide the back-end specifications, i.e.  the type of the
%spectrometer (column~7), the spectrometer bandwidth (column~8) and the
%spectral resolution (column~9).  
Please note that all back-ends can usually be
attached to all front-ends and are thus given in just one line of the
table.  
%The last two columns contain the possible observing methods,
%as discussed in Sec.~\ref{observing-rt}, and the half power beam width
%in each frequency range of the receiver.
 
Nearly all radiotelescopes use SIS mixers that provide low receiver
noise temperatures and work in DSB mode, in which emission from
the upper and lower sidebands are overlaid and indiscriminately
detected. The receivers have large tuning ranges although for some
RTs, certain frequency windows are intentionally excluded (e.g.  the
280--320 GHz or 380--420\,GHz range) since the atmosphere is opaque at
these frequencies due to strong H$_2$O and O$_2$ line absorption.  As
back-ends, acousto-optical spectrometers, filterbanks, and  
recently Fast Fourier Transform Spectrometers are widely
distributed and offer a wide range of bandwidths and spectral
resolutions.  The receiver bandwidth is typically around 1\,GHz.

%-------------------------------------------------------------------------
\section{Using radiotelescopes for atmospheric line observations}  \label{discussion} 
%
%-- {\bf TO DO}: Very short summary of results (promising !) and outlook how RT can be used 
%   in future: detection of weak species, long-term monitoring, exploiting 
%   new sites (in parallel to the existing ones like JCMT, CSO, IRAM..) and 
%   thinking of APEX, ALMA..  Which species are interesting to detect/monitor ? 
%   Link Earth other planets. 
%Nicolas: In your "TO DO" paragraph, you were mentioning "Link Earth to other planets". 
%Well, for the HCN business, of course one important thing is the oxidizing capacity of the 
%Earth atmosphere compare to the other planets (reducing atmosphere).
%
The HCN observations presented in Paper-II, performed using three
well-known sub-mm radiotelescopes, together with a long tradition to
use RTs for sporadic measurements of atmospheric lines raise the
question how useful RTs are for such measurements.  In particular with
respect to new antennas (ALMA, LMT) and envisaged projects (DOME-A/C,
CCAT) it is of great interest to explore the possible use of these
telescopes for observing telluric lines.  Accompanying the exploitation 
of these very good observing sites, the technical development of
front- and back-ends opens now also the possibility to observe up to a
frequency of 1.5\,THz using broad band (bandwidth $>$ 2\,GHz)
receivers and spectrometers.  This would enable to observe a number of
interesting molecular and possibly atomic species at high-frequencies
covering the stratosphere and mesosphere.
%
%{\bf Nicola: Which ones?  CH3CN?  HCN?  CO?  ....  to which extend
%  are1 these species excited (JO: All molecular transitions are in
%  LTE!)  and thus observable (results of forward  models!)?  
%  What do  we learn from them?}\\
%{\bf JO: Not so simple - requires a detailed study what is
%  theoretically measurable.  One could plot the three last columns of
%  the moliere spectral line file
%  \begin{verbatim}
%  sens_0000_3000.ul-0,1K.verdandi_208.val
%  \end{verbatim}
%  for the up-looking  geometry which provide estimated line  strengths
%  for each spectral line in  Moliere/5 with Tb$>$0.1K.}\\
%
Due to the high sensitivity and angular resolution, it is now also
feasible to investigate the atmosphere of other planets (e.g.
detection of CO and other molecules in the Martian atmosphere, see
\citet{encrenaz2004} for a summary).

%The exploration of the Earth's atmosphere is also of great importance 
%for understanding generally 

In the following, we summarize advantages and disadvantages of using
RTs for atmospheric line observations: \\

\paragraph*{Advantages}
\begin{itemize} 
\item{}  RTs     are   installed on   explored    sites  that  offer 
  well-developed logistics and simple access (exceptions are 
  DOME-A, C and F in Antarctica).  The   geographical  coverage is large
  though  only a few instruments exist  at high latitudes.  The use of
  RTs is normally  open to the  aeronomy community and observing  time
  can be granted via proposals or guaranteed time.
\item{} Some   telescopes have  spare   time that   is  not  used  for
  astronomical  observations  and   could be  allocated   for aeronomy
  measurements.  Examples are the JCMT during daytime or Mopra and
  KOSMA during the summer period.
\item{} RTs are  flexible measurement systems providing front-ends and
  back-ends suitable for a  wide  range of applications, ranging  from
  observations   of strong narrow  lines that  require a high spectral
  resolution to measurements of weak broad  features (rare and complex
  molecules in  the atmosphere, observations  of external galaxies and
  the  Galactic center).  Of particular  interest for  aeronomy is the
  detection  of weak and  pressure-broadened   lines which is feasible 
  by the provided   technical equipment. Nearly all RT
  receivers  are  of SIS type, providing  very  low noise temperatures
  (typically $<$100\,K\,@\,200\,GHz). New instruments  using 
  HEBs (Hot Electron Bolometer) extend the accessible frequency range 
  up to $\sim$1--2 THz.
\item{} RTs   have implemented  robust  and well  tested   methods for
  precise  calibration  to  obtain     correct main beam    brightness
  temperatures  (see Sec.~\ref{observing-rt}). The  
  calibration error, however, can be of the order of 5--10\%, 
  depending on site and weather conditions. Standard procedures to
  determine the    tropospheric  opacity   are   also   provided  (see
  Sec.~\ref{tropospheric_correction}).  As  an  additional    feature,
  astronomical sources of well-known emission strength  
  can be   used in two  ways for   improving the
  calibration (for example the sideband-ratio): (i)~the astronomical
  object   is observed {\sl  before}  and  {\sl after} the aeronomical
  observation or (ii)~the astronomical object is  all the time visible
  in the spectrum. In  the latter case,  blending  of lines has  to be
  avoided.
\item{} RTs offer possibilities  to overcome baseline problems  due to
  standing   waves  and  reflections  by  moving   the subreflector by
  $\lambda$/4
%   ($\lambda$ is the  observing  wavelength) 
or by random moving. 
\item{} RTs are steerable instruments and can thus be used for 
  skydips and/or scanning the atmosphere.
\item{} Standard software tools are usually available for RT operation
  and calibration. 
\end{itemize} 

\paragraph*{Disadvantages}
\begin{itemize} 
\item{}   Since the instruments   are  not  movable, the  geographical
  location of  RTs is restricted to certain  geographical areas on the
  globe.
  % More flexibility  is   thus only  provided by  sondes  and  aircrafts.
\item{} Observing time --\,in particular for the sub-mm telescopes and
  for a sophisticated instrument  like ALMA\,--   is  very valuable   and
  sometimes difficult to  obtain.   Long-term monitoring projects  are
  thus hard to  realize unless spare time (e.g.   daytime at JCMT,  see
  above) is efficiently used.  This, however, may impose a bias on the
  observations  depending on  the scientific application. For   example, 
  it may be difficult  to
  observe diurnal and seasonal variations of some molecules as well as
  to analyse their long-term variability.
\item{} RTs are  not  concepted for atmospheric measurements  so  that
  special  observing techniques  (e.g.  subreflector movement  for the
  reduction of  standing   waves) are often not implemented   in the normal
  procedures.   However,  astronomical  observations may  benefit from
  optimisations and standard  implementation  of these   techniques as
  well.
\item{} A calibration uncertainty of up to 10\% can be too high for 
 certain aeronomical applications that demand a high precision. Pure 
 detection projects or monitoring of strong lines (for example mesospheric CO) 
 are not problematic.  
\end{itemize} 

\smallskip
Considering the pro and contra arguments for the use of
radiotelescopes for atmospheric line observations, we come to the
conclusion that RTs offer indeed a very good opportunity for certain
applications in aeronomy.  In particular exploratory observations of
'exotic' and thus probably weak lines or high-frequency transitions of
standard molecules can easily be performed with RTs before a dedicated
aeronomy receiver is built and installed for long-term monitoring of
trends and variability.  It should, however, be considered that the
atmospheric abundance of some molecules of interest for aeronomy is
high enough only at high latitudes where currently no (sub)-mm 
telescope is installed.  Long-term monitoring projects of
standard molecules with strong lines --\,like e.g.  the observation of
strato-mesospheric CO\,-- are easy to do with standard RT observing
modes and have in fact already been realized \citep[e.g.][7-year time series 
of atmospheric CO using KOSMA]{boes1994}.
%-------------------------------------------------------------------------
%-------------------------------------------------------------------------
\section{Prediction of the Earth's atmospheric transmission}  
\subsection{Introduction}
The simulation of the atmospheric transmission is essential for all
astronomy and middle atmospheric applications. In {\sl astronomy}, the
knowledge of atmospheric transmission at a given frequency is
mandatory to asses the quality of an existing or future observatory
site.  It is also essential for the calibration since it is  
required to calculate the system noise temperature
in case the opacity is not independently determined.  {\sl Remote
observations} of the Earth's middle atmosphere require to model
accurately the observed emission spectra. 'Forward models' are used to
describe the radiative transfer, spectroscopy, and instrument
characteristics and to compute weighting functions with respect to the
searched atmospheric quantities.  A versatile forward and inversion
model for the (sub)-mm wavelengths range
used in many aeronomy applications is the MOLIERE-5 code ({\em
Microwave Observation LIne Estimation and REtrieval}).  See
\citet[]{schneider2003, urban2004a, kasai2006} for its use for the
analysis of ground-based and airborne microwave observations, and
\citet[]{urban2005b, urban2007a} for applications with respect to
space-borne missions.  MOLIERE-5 has also been used for
prediction/feasibility studies of future satellite projects for the
exploration of the Earth and Mars atmospheres \citep[]{urban2000a,
urban2005a}.
%,[Kerridge et al. 2005].

There are other atmospheric models that have been used to date (see
\citet{pardo2001} for a short summary).  A discussion of these models
and a comparison with MOLIERE-5 is the subject of Paper-III (Schneider
et al.  2009, in prep.).  Here we focus on a short description of
MOLIERE-5 and present the results of the modeled atmospheric
transmission in the sub-mm/FIR range for different observatory sites.

\subsection{The forward model MOLIERE-5} \label{moliere}

% GEBNERAL
A detailed mathematical description of the MOLIERE-5 forward and
inversion model and the underlying principles is provided by
\citet{urban2004a}.  Designed for a variety of applications, the 
model comprises modules for spectroscopy,
radiative transfer, and instrument characteristics.  Important
features of the absorption coefficient module are the line-by-line
calculation as well as the implemented H$_2$O, O$_2$, N$_2$, and
CO$_2$ continuum models \citep{borysow1986, ckd1989, liebe1993,
rosenkranz1993, pardo2001}.  Note that MOLIERE-5 is a 'clear-sky' 
model that simulates atmospheric gas-phase emission and absorption. It
does not include the radiative effect of cloud droplets or
ice (see, e.g., \citet{prigent2006} for a discussion of this effect).

% Radiative transfer 
The radiative transfer module allows for calculations in different
geometries such as limb and nadir sounding from orbiting platforms as
well as up-looking observations of ground-based or airborne sensors.
A spherically stratified (1-D) emitting and absorbing (non-scattering)
atmosphere in local thermodynamical equilibrium is assumed, i.e.  the
source function is given by Planck's function.  The geometrical
radiation path is corrected for the effect of refraction.  Weighting
functions, required for inversions, are calculated by differentiating
the radiative transfer equation analytically after discretisation.
The radiative transfer model is supplemented by a sensitivity module
for estimating the contribution to the spectrum of each catalogue line
at its centre frequency, enabling the model to effectively filter
large spectral data bases for relevant spectral lines.

The spectroscopic parameters used in MOLIERE-5 are obtained from the
VERDANDI catalog \citep{eriksson1999} that contains the JPL
\citep{pickett1998} and HITRAN \citep{rothman1998} line parameters up
to 10 THz. The catalog is updated regularly and available at
http://www.chalmers.se/rss/.  We included all spectral lines in this
catalog of molecules with a contribution of 0.1 K brightness
temperature to the spectrum up to 10 THz.

Several independent modules permit accurate simulation of instrument
characteristics such as the antenna field-of-view, the sideband
response of a heterodyne receiver, as well as the spectrometer
bandwidth and resolution.  Frequency switched observations may also be
modelled. These features, however, were not used for the
forward-calculations presented here but are required to model real
observations. This particular feature makes MOLIERE-5 a very useful
tool for aeronomy and astronomy applications and makes it different
from other atmospheric models where instrumental effects are not -- or
only partly included.

We use the following H$_2$O line and continuum and dry air models for
the calculations presented in this paper. \\

\noindent {\sl Frequency range 1 -- 1000 GHz} 

The absorption due to water vapor is modeled as described in
\citet{pardo2001}: all H$_2$O lines up to 10 THz are calculated in
order to account for the H$_2$O far-wing absorption below 2 THz. In
addition, a pseudocontinuum water vapor absorption ($\propto \nu^2$)
is derived from Fourier Transform Spectrometer (FTS) obervations
in the 0.5--1.6 THz regime \citep{pardo2005}.  The absorption
due to dry air continua is (i) collisionally induced absorption
(N$_2$ and O$_2$) that is also contained in the observationally based
model of \citet{pardo2001}, and (ii) relaxation (Debye) absorption of O$_2$
that is implemented using the formalism described in \citet{rosenkranz1993}
and \citet{pardo2001}. \\

\noindent {\sl Frequency range 1000 -- 2000 GHz} 

Based on FTS measurements in the 0.5--1.6 THz range, \citet{pardo2005}
showed that the $\propto \nu^2$ approximation for the H$_2$O
pseudocontinuum starts to fail for frequencies $>$1 THz and that the
exponent seems to be lower than 2 for $\nu >$1.1 THz.  We thus use
the semi-empirical MT-CKD-2004 model (http://www.rtweb.aer.com,
\citet{clough2005}), a continuation of the older CKD-model
\citep{ckd1989}.  It includes continuum absorption due to water vapor,
nitrogen, oxygen, and CO$_2$.  A new feature is that the self- and
foreign continuum models are each based on the contributions from a
collision induced component and a line shape component.  These are
applied to all water vapor lines from the millimetre regime to the
visible, and the results are summed to obtain self and foreign
continuum coefficients from 0--20000 cm$^{-1}$ (0--600 THz).
%
%the pseudocontinuum 
%absorption for collisions of H$_2$O with O$_2$ and N$_2$ as it is implemented 
%in the MT-CKD model (http://www.rtweb.aer.com). 
%by taking into account the far wings of all water vapor lines. 
%
The high-frequency collision-induced spectrum of nitrogen is extracted
from the formulation of \citet{borysow1986}. The model is scaled with
a factor 1.34 as suggested by \citet{boissoles2003}.
%
%for the air-air continuum: Boissoles J., Boulet C., Tipping R. H.,
%Brown A. and Ma Q. , Theoretical calculation of the translation-rotation
%collision-induced absorption in N2-N2, O2-O2 and N2-O2 pairs, Journal
%of Quantitative Spectroscopy & Radiative Transfer, 1-4, 505-516, 2003.
%
%A more detailed discussion and comparisons of the 
%various continuum terms will be presented in Paper-III. 

\subsection{Contribution of dry and wet component}

\begin{figure*}
  \begin{center}
  \includegraphics[width=8cm,angle=-90]{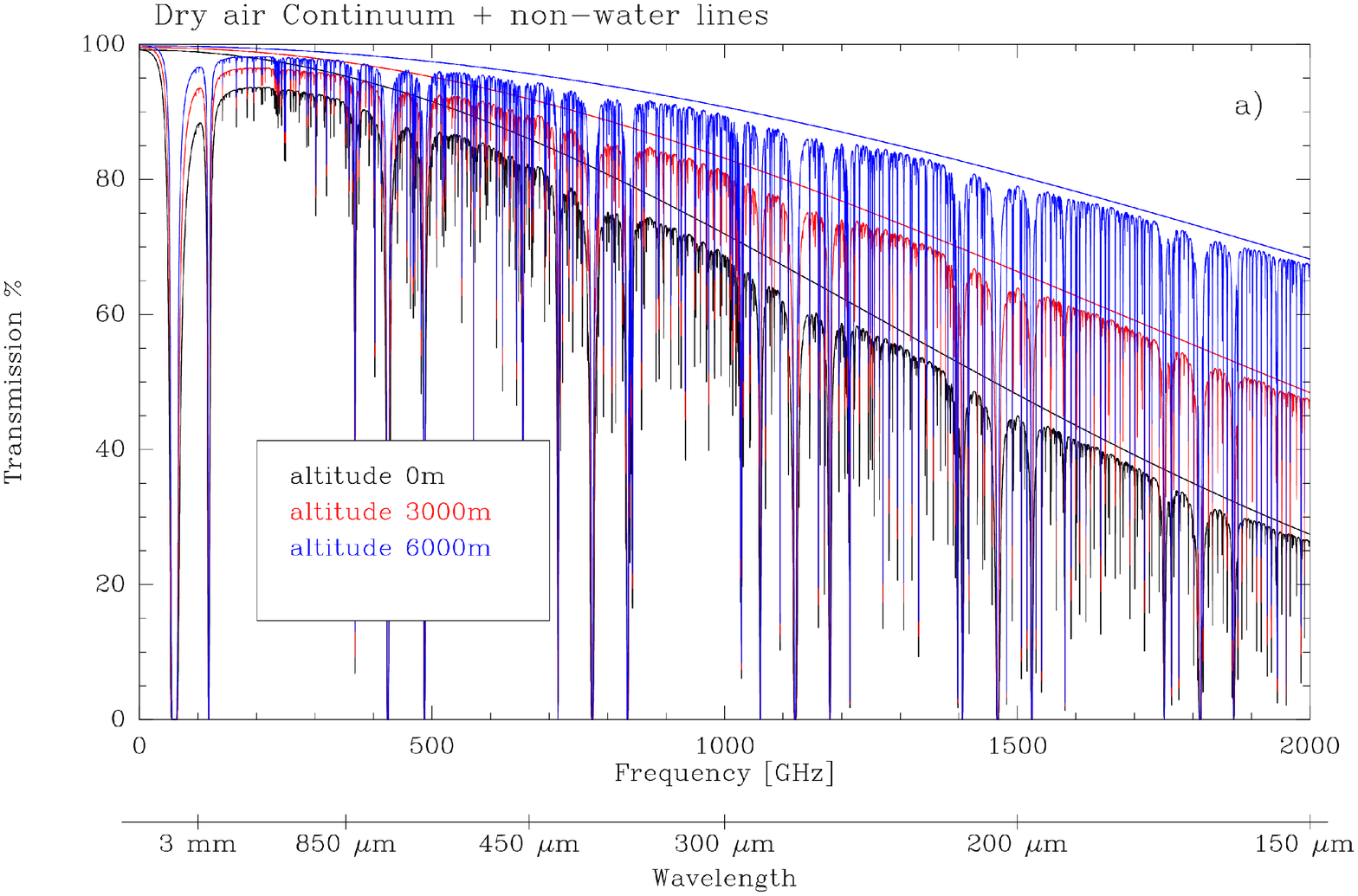}
  \includegraphics[width=8cm,angle=-90]{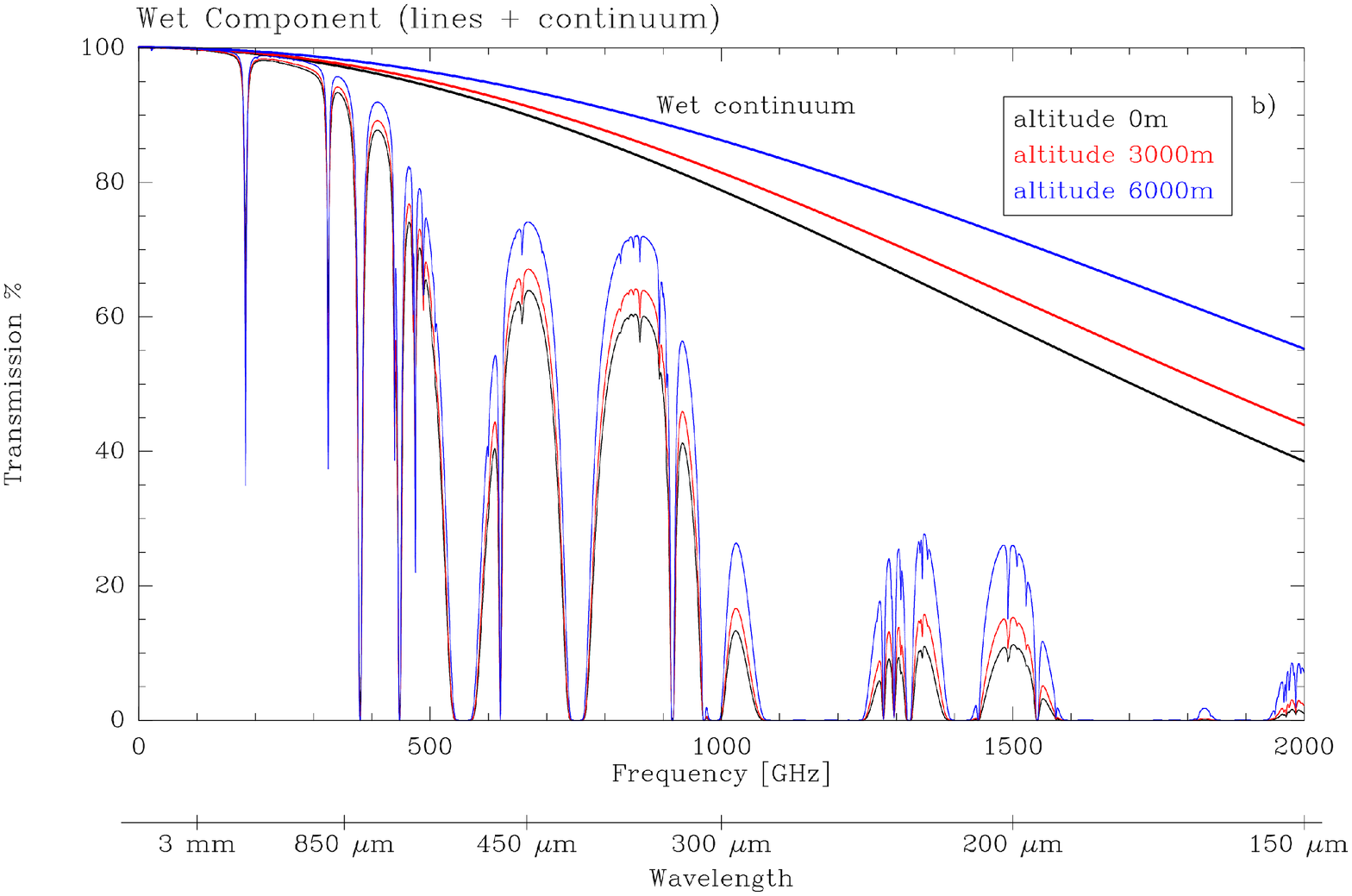}
  \end{center}
  \caption
{{\bf a)} Transmission calculated for three different altitudes,
   considering only the dry air component. The continuous curves show
   collisionally induced absorption (N$_2$ and O$_2$) from the models
   of \citet{pardo2001}, and relaxation (Debye) absorption of O$_2$
   (\citet{rosenkranz1993}, \citet{pardo2001}).  The other curves
   contain additional absorption due to all lines except water
   (including some strong lines like O$_2$ or O$_3$).  {\bf b)}
   Transmission calculated for three different altitudes, {\bf only
   taking into account} the wet component (line and continuum), based
   on the model of \citet{pardo2001}. A common pwv of 0.3mm is
   used. The water continuum alone is included for comparison.  The
   temperature and pressure profiles {\bf for all curves} are from the
   45$^\circ$ N U.S. standard atmosphere, the spectral resolution is
   400 MHz (dry air) and 200 MHz (wet component), respectively.}
   \label{air}
\end{figure*}

In order to to characterize the different contributions of the
transmission, we separated the dry and wet component and performed
model calculations for three different altitudes (0, 3000 and 6000m).
We used a standard 45$^\circ$ N H$_2$O-profile (scaled depending on
altitude for the wet component to give an integrated column of
0.3 mm of pwv for all altitudes) and a 45$^\circ$ N temperature and
pressure profile.  Figure~\ref{air}a shows the purely dry air
continuum (Pardo-2001 collisional dry air (N$_2$-O$_2$) and
O$_2$-Debye continua) alone and curves including all non-water
lines. (For simplification, we did not separate the different continua
depending on frequency as explained in Sec.~\ref{moliere}.) Strong
lines from O$_2$ or O$_3$ cause significant broad-band absorption, in
particular at lower frequencies.  The dry air continuum (without
O$_2$ and O$_3$ lines) is dominated by N$_2$+O$_2$ non-resonant
absorption below 300 GHz and only by N$_2$ non-resonant absorption
above 300 GHz.

%The non-resonant nitrogen absorption accounts for the characteristical
%decrease of the transmission (increased absorption) towards higher
%frequencies.

Figure~\ref{air}b displays the wet component alone (Pardo-2001
H$_2$O-continuum and all water lines) at a pwv of 0.3mm.  The water
vapor spectrum is characterized by strong and broad water absorption
lines and an increasing continuum absorption with frequency. Though
this plot shows only the wet component at 0.3 mm (which is a rather
low value, found only at the driest sites on Earth), it is obvious
that the absorption due to the wet component -- in particular H$_2$O
line absorption -- is generally more important than the dry component
(lines+continuum). Suitability of a given site for astronomical or
atmospheric observations depends largely on the specific (local)
weather conditions at the site, including humidity and
cloudiness. Here we can only generally state that low humidity and
high altitude are both beneficial for the observation conditions,
whilst in practise also logistical issues such as accessibility are of
major relevance.

% Martina's Kommentar (den ich immer noch nicht verstehe aber auf dem 
%man antworten muss):  
%
%a discussion of the composition of the transmission,
%e.g. why does the same amount of PWV leads to different amounts of
%transmission for telescopes at different altitude? I would find plot
%that shows the transmission due to the dry component alone at
%altitudes of e.g. 0, 3000 and 6000m as well as a second plot which
%shows the line broadening of the wet component alone e.g. 1mm PWV at
%0, 3000, and 6000m very interesting. (This would answer e.g. the
%question whether for the best transmission at a certain frequency a
%dry site would be sufficient or whether the site needs to be at a
%minimum altitude as well.)  

\subsection{Assumptions of the model and real atmosphere} \label{limit} 

Our model calculations are based on the clear sky assumption, i.e
do not include the effect of water clouds and cirrus.  Absorption of
thin and sub-visual cirrus is typically negligible to radio
observations, but the observer should watch out for 
significant ice and more importantly liquid water clouds which affect the
observations or calibration procedures.

%For astronomical observations, {\sl clouds} strongly disturb the
%observations because they do not form a homogeneous 'background' (like
%cirrus does) but vary temporarily and spatially (on very short
%timescales of a few seconds).  Astronomical/aeronomical integration
%times are typically much longer, so that even if individual short
%observations in frequency-swichting or fast beam-switching mode are
%possible, the integration of many short scans leads to an additional
%calibration uncertainty.}

Another aspect to consider is the timescale for the stability of
the transmission. Even under 'clear-sky' conditions, a variation of
the attenuation of the atmosphere due to changes of humidity and 
temperature causes slow changes in transmission
(timescale of several hours) and thus creates difficulties in
calibration. This can be corrected for by regularly performing
sky-dips.  Moving inhomogeneities of the water vapor distribution,
however, cause rapid fluctuations (seconds to minutes) in the
atmospheric radiation, leading to noise in the observations (called
'seeing' in the optical wavelength range and 'sky noise' in the
(sub)mm-range) which are not easy to correct for. 

\subsection{Specific input parameters} \label{param} 

Since tropospheric water is the main atmospheric absorber at
(sub)-mm wavelengths and varies significantly with time, we need to
provide a common base for inter-comparison with results from other
models. We thus present model calculations for different amounts of
precipitable water vapor (pwv).
%
%We used adaquate temperature- and pressure profiles (see subsections) for
%each site  and  the  U.S.   standard  H$_2$O  profile (reference)  for
%45$^\circ$N.  
%
Taking into account the altitude of the site, we produced different H$_2$O
vertical profiles by scaling the tropospheric part of the
U.S. standard profile accordingly (i.e. so that the integrated H$_2$O
column above the site corresponds to a certain pwv). 

%The typical tropospheric scale height of H$_2$O is 2\,km.

For O$_2$ and O$_3$, H$_2$O, and the minor species (e.g. NO, N$_2$O,
CO, SO$_2$, HNO$_3$, etc.)
%OH, HCl, ClO, OCS, H$_2$CO, HCN, CH$_3$Cl, H$_2$O$_2$, ClNO$_3$, 
we use typical mid-latitude vertical profiles.  
%Profiles of HOCl, HO$_2$, BrO, HOBr, OClO, Cl$_2$O$_2$ are based on models
%(SLIMCAT,\citep{chipperfield1996}) and those of HBr, OBrO, CH$_3$CN on
%observations. 
%
The full set of profiles as well as Temperature (T) and
pressure (p) profiles can be found on our website. The T,p-profiles
used here are climatological monthly averages (July for ALMA and
DOME-C, January for KOSMA and JCMT/CSO) and zonal means over 10
degrees. For example, the site of ALMA at 23$^\circ$S was obtained by
averaging T,p-profiles between 20 and 30 degrees South.

%Molekuele in radio_45n.vmr_list
%
% H2O -> prozentual veraendert je nach pwv
% 45N profile: 
% O3, N2O, CO, O2,NO, SO2, HNO3, OH, HCl, ClO, OCS, H2CO, HCN, CH3Cl   
% H2O2, ClNO3, H2O-16/17/19, O3-16/17/18/asym 
% Slimcat-Modelle:
% HOCl, HO2, BrO, HOBr, OClO, Cl2O2 
% Observed: 
% HBr, OBrO, CH3CN,

%are obtained from the
%%ODIN/SMR \thanksref{odin} \thanks[odin]{Odin is a Swedish-led
%satellite project funded jointly by the Swedish National Space Board,
%the Canadian Space Agency, the National Technology Agency of Finland,
%and the Centre National d'Etudes Spatial in France. The Swedish Space
%Corporation has been the industrial prime contractor.}  climatology
%(see http://diamond.rss.chalmers.se/odin for information about this
%satellite that made atmospheric and astronomy observations in the
%submm wavelength range).  The data is available on the french database
%'ETHER' (http://ether.ipsl.jussieu.fr). 

All models shown in this paper were calculated for 90$^\circ$
elevation, i.e. zenith view, and have a frequency resolution of
200\,MHz.  A large number of additional computations at higher
spectral resolution, different water vapor columns as well as zooms
into particular frequency ranges of potential interest can be found on
our website. There, we also provide graphical displays and numerical
data files (2-column ascii-format with frequency and transmission) for
more cases, e.g., other Antarctic stations, CCAT, the NASA/DLR
aircraft SOFIA (Stratospheric Observatory for Infrared Astronomy), and 
balloon experiments.

%{\bf Jo:  Hier kurz zusammenfassen  was fuer die Rechnungen angenommen
%  wurde:    Spectral  lines  (of    all    molecules with  significant
%  contribution to the  spectrum) up to 3THz,  H2O continuum  model [],
%  N2-N2 continuum model [], O2-O2 and O2-N2 continuum model [], zenith
%  view     (90deg  elevation),  species    climatology   (from Odin?),
%  temperature/pressure from ?} 

%{\bf  Nicola:      Im     folgenden    Diskussion     der    einzelnen
%  Transmissionskurven. Erklaerung welche T,p-Profile eingehen.  Kurzer
%  Vergleich mit existierenden Modellen ???? Im Internet sind fuer alle
%  sites  plots zu finden.  Woher  koennen die Unterschiede kommen?  Wo
%  liegen  die Grenzen  von  MOLIERE  (das  koennte eigene   subsection
%  werden. Wie verlaesslich ist der Frequenzbereich hoeher als 1 THz.}

%{\bf Nicola: Man  koennte  auch Transmissionskurven fuer  verschiedene
%  Eingabeparaemeter berechnen ?  } {\bf JO: Das  ist am besten - dann
%  die Vergleichsergebnisse diskutieren (verschiedene pwv, verschiedene
%  Kontinuummodelle,  etc.).     Dann braucht  man  hiere   auch keinen
%  Vergleich mit ATM.}

\subsection{Transmission for selected sites}

\begin{figure*}
  \begin{center}
  \includegraphics[width=9cm,angle=-90]{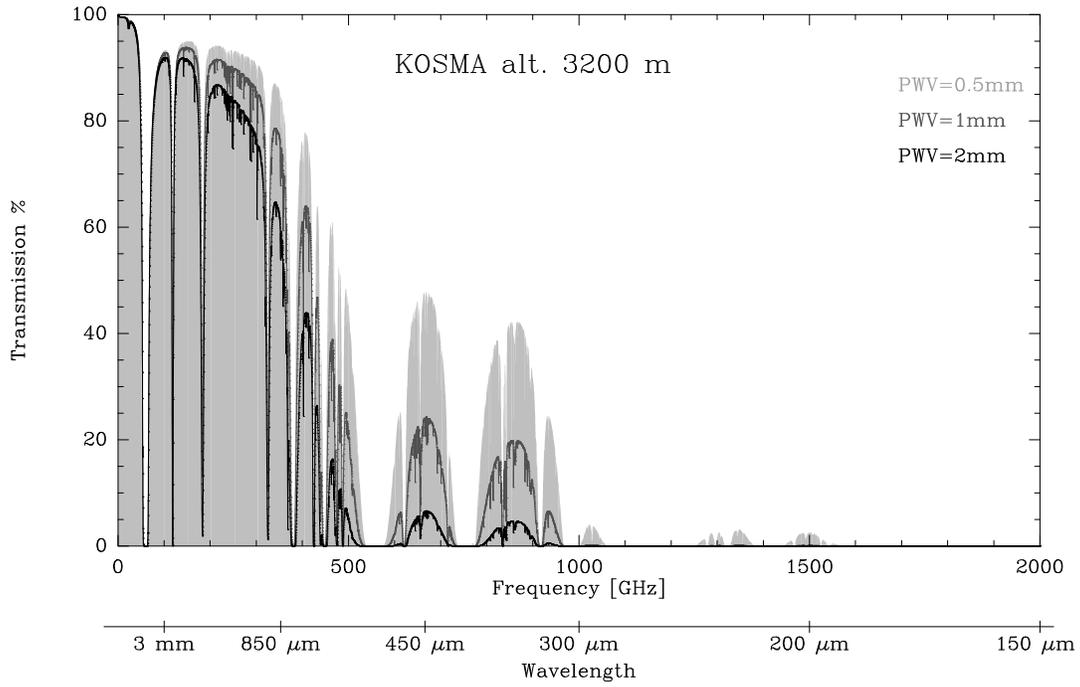}
  \end{center}
  \caption{Transmission curves (in \%) for 0.5, 1, and 2\,mm
    precipitable water (pwv) for the KOSMA site
    (Gornergrat/Switzerland).}
   \label{kosma}
\end{figure*}

\begin{figure*}
  \begin{center}
  \includegraphics[width=9cm,angle=-90]{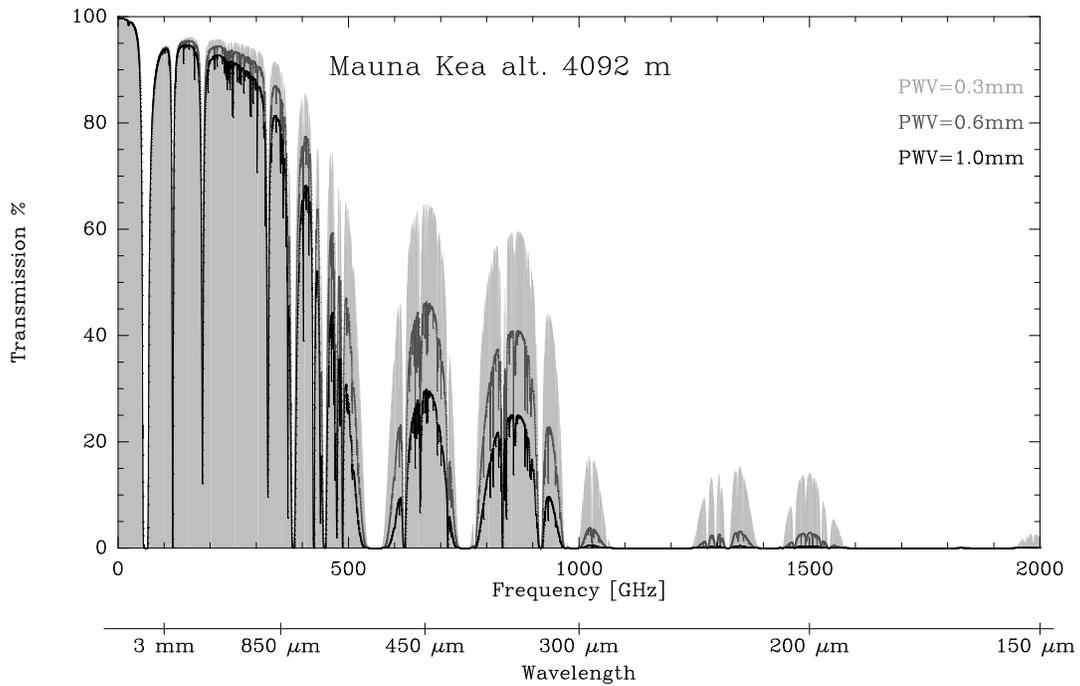}
  \end{center}
  \caption{Transmission curves (in \%) for 0.3, 0.6, and 1\,mm
    precipitable water (pwv) for the JCMT/CSO site (Mauna
    Kea/Hawaii).}
   \label{jcmt}
\end{figure*}

\begin{figure*}
  \begin{center}
   \includegraphics[width=9cm,angle=-90]{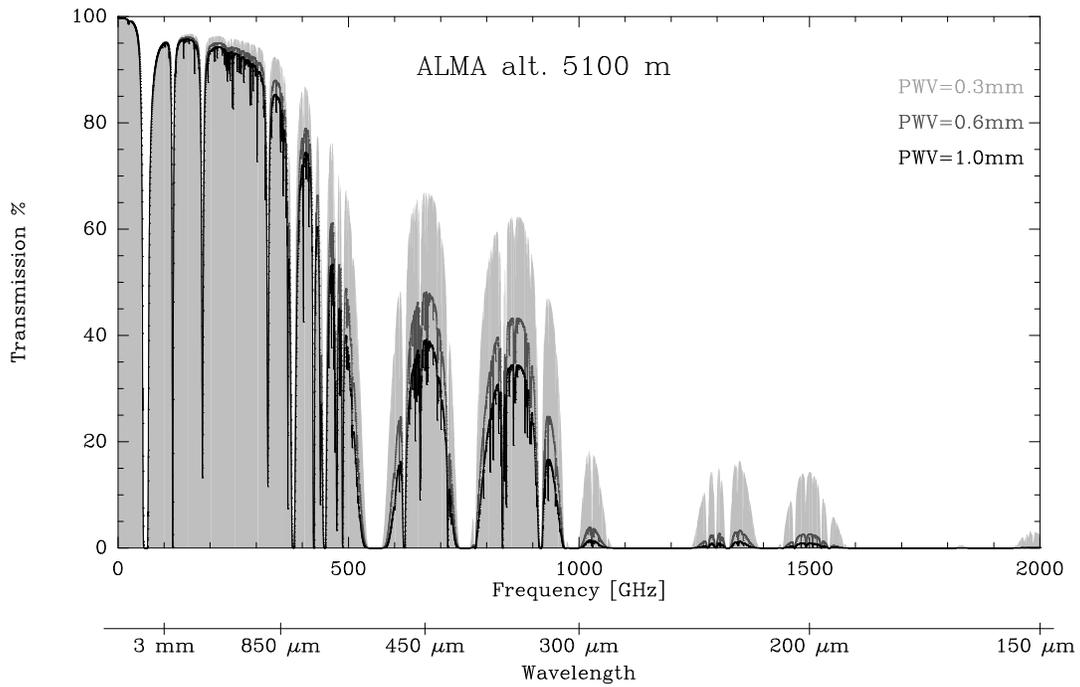}
  \end{center}
  \caption{Atmospheric transmission (in \%) for 0.3, 0.6, and 1\,mm
    precipitable water (pwv) for the
    ALMA/APEX site (Llano de Chajnantor).}
  \label{alma}
\end{figure*}

\begin{figure*}
  \begin{center}
   \includegraphics[width=9cm,angle=-90]{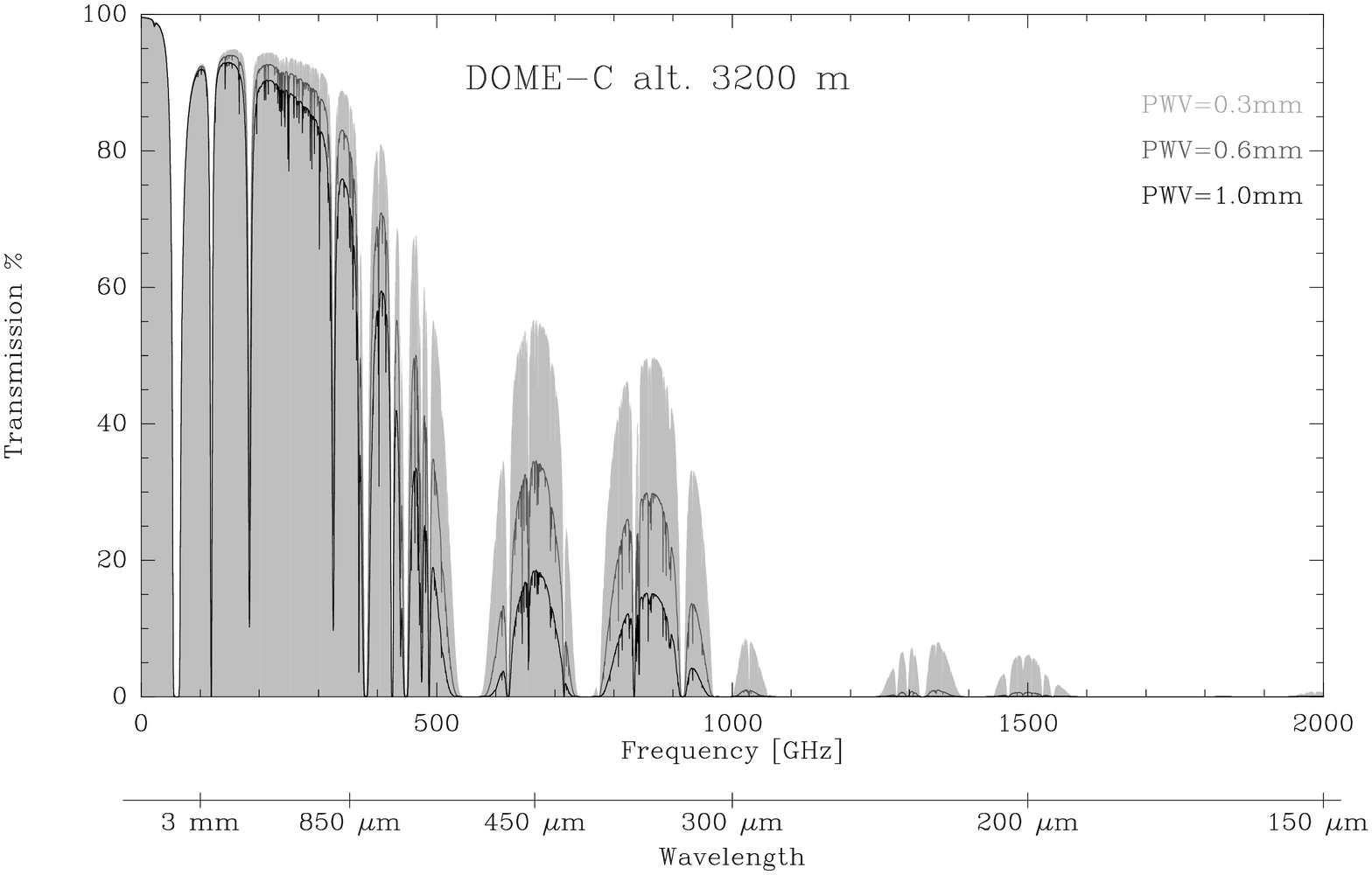}
  \end{center}
  \caption{Atmospheric transmission (in \%) for 0.3, 0.6, and 1\,mm
    precipitable water (pwv) for the DOME-C site (Concordia
    station/Antarctica).}
  \label{domec}
\end{figure*}

\begin{figure*}
  \begin{center}
   \includegraphics[width=9cm,angle=-90]{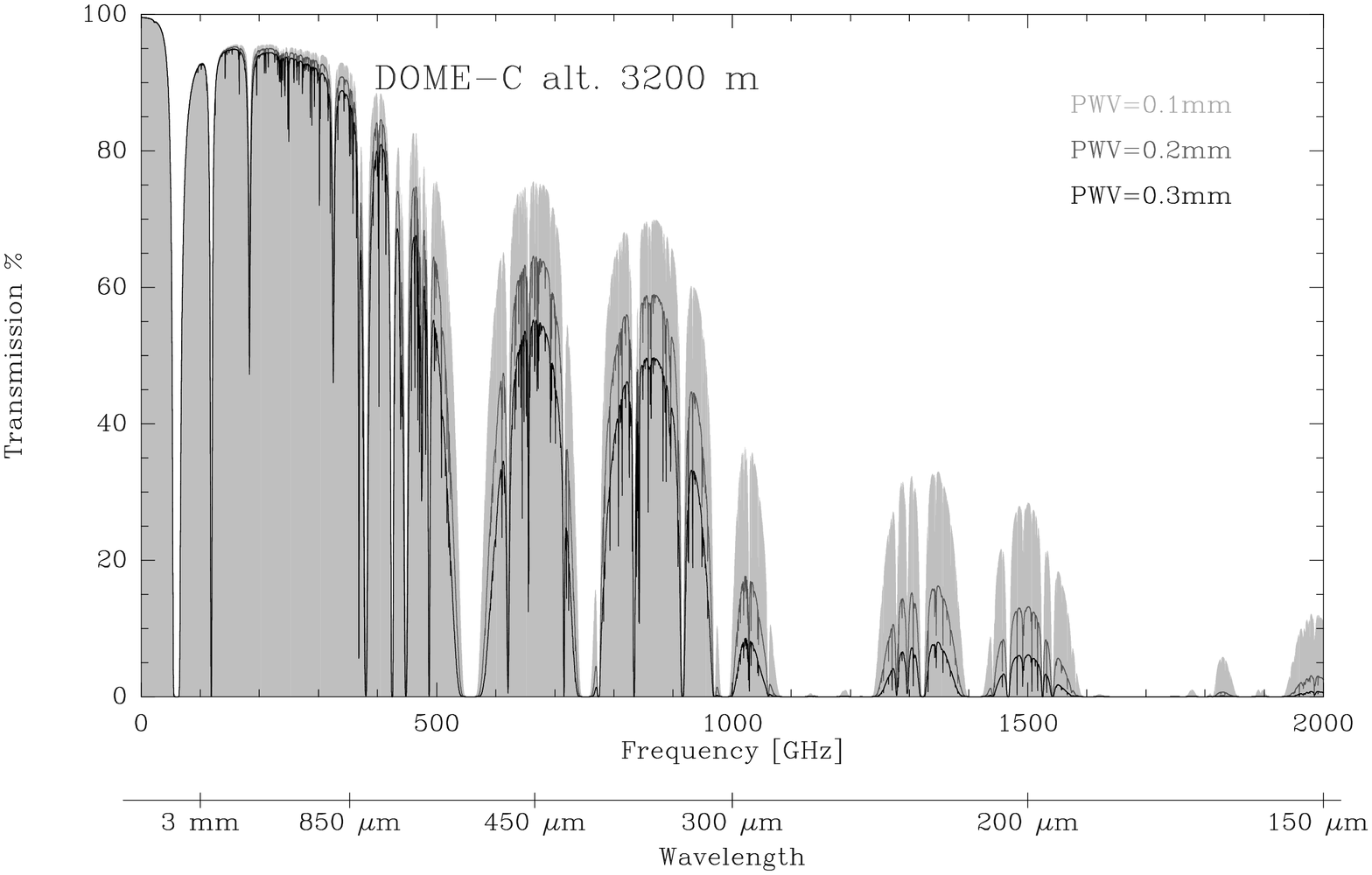}
  \end{center}
  \caption{Atmospheric transmission (in \%) for 0.1, 0.2, and 0.3\,mm
    precipitable water (pwv) for the DOME-C site (Concordia
    station/Antarctica).}
  \label{domec1}
\end{figure*}

Figures \ref{kosma} to \ref{domec1} show the results of our forward
calculations in the frequency range 0--2000\,GHz (frequency resolution
200 MHz) for four different observatories.  We selected the KOSMA
(Gornergrat, Swiss Alps) and CSO/JCMT (Mauna Kea) sites because
atmospheric HCN observations with these telescopes will be presented
in Paper-II.  The ALMA/APEX (Chajnantor) site is of large importance
for astronomy and represents a very high-altitude ($\sim$5000\,m) site
at a latitude of about 23$^\circ$S.  DOME-C is a potentially very
interesting location for performing aeronomy observations related to
polar ozone chemistry since it is located inside the Antarctic polar
vortex at 75$^\circ$S,123$^\circ$W (Concordia station at 3260m
altitude, see http://www.concordiastation.org). Calculations were done
using three different amounts of precipitable water vapor, depending
on site and published literature. We thus chose for KOSMA a pwv of
0.5, 1, and 2mm while Mauna Kea, DOME-C, and ALMA show smaller pwv's
on average so that we run models with pwv=0.3, 0.6, and 1mm. We want
to emphasize that in this study, we only present transmission curves
and do not discuss the general 'quality' of different sites. For that,
statistics on the cumulative probability for the fraction of time
above a certain transmission value have to be considered as well as
other meterological factors like atmospheric stability, wind
etc.. See, e.g., \citet{lawrence2004} and Minier et al. [2009 in
prep.] for DOME-C,\citet{marks2002} for DOME-A, and
\citet{oterola2005} for ALMA.

\subsubsection{KOSMA (3200m)} 

It becomes obvious from Fig.~\ref{kosma} -- and is known since long --
that the millimetre wavelengths range (up to 400 GHz) has the best
transmission with values better than 40\% even for a pwv of 2mm.
However, a few strong and broad absorption lines of H$_2$O, O$_2$, and
O$_3$, are present in this frequency range. These are for example
O$_2$ lines at 60 and 119 GHz and the H$_2$O line at 183 GHz. 
%, 325, 380, 489,557, 620, and 750 GHz.  
Since the atmosphere is very transparent in
this frequency range for all sites discussed in this paper, we will
not go into more details here but focus on the higher frequency ranges
($>$400 GHz). Only a few atmospheric windows open at $\sim$410, 480,
$\sim$660, and 850 GHz where a transmission of up to 40\% can be reached. The
THz-range, however, has only few windows: one at 1050 GHz, one 
at 1300 GHz, and one at 1500 GHz with a transmission rarely above 10\%. 
Successfull astronomical observations at KOSMA were performed at
frequencies of around 115, 230, 345, 461, 492, and 806/809 GHz and
thus demonstrate the capability of this site as a sub-mm 
observatory. Further information on the atmospheric transparency at
KOSMA can be found at http://ph1.uni-koeln.de/kosma and in
\citep{kramer1990}. 

%summer: 2-12mm, 20\% of time with pwv$<$1.7mm 
  
%Compare with results shown on website (1 mm) 
%http://ph1.uni-koeln.de/kosma 
%or technical report (1, 2mm)  
%http://hera.ph1.uni-koeln.de/~kramer/publications.html techmem_wetter.ps.gz

\subsubsection{JCMT/CSO (4100m)} 

Transmission curves for 0.3, 0.6, and 1mm pwv are displayed in
Fig.~\ref{jcmt}. Although the higher altitude of Mauna Kea compared to
the KOSMA site in the Swiss Alps increases the transparency of the
atmosphere, this effect is only significant at high frequencies.  The
windows at 660 and 850 GHz have a transmission of 30\% for pwv=1mm at
Mauna Kea while at KOSMA, the transmission at the same pwv is 5\%
less.  In contrast to KOSMA, the high-frequency windows now open up
with transmissions up to 20\% and thus allow occasionally observations
of atmospheric tracers as well as astronomically interesting lines up
to 950 GHz (e.g., NH$_2$ at 902 GHz). Experimental observations above
1 THz become feasible (e.g., NH$^+$ at 1.0126 THz, p-H$_2$D$^+$ at
1.37 THz, [NII] at 1.46 GHz. Observable lines for aeronomy are for
example CO, NO, NO$_2$, OH, HO$_2$, H$_2$CO, HOCl, HOBr, HCN.  Some
other trace gases like N$_2$O, OCS, and CH$_3$Cl have their maximum
emission below 1 THz. The reason for the better transmission is
the higher altitude of the site (3200m for KOSMA and 4100m for Mauna
Kea) and the generally (slightly) lower pwv.  A comparison with a
transmission calculator (see
http://www.submm.caltech.edu/cso/weather/atplot.shtml) provided on the
CSO website (based on the ATM model of \citet{pardo2001}) shows a
deviation to ATM of less than 5\% for all frequency ranges and values
of pwvs. This is an interesting result, in view of the different
implementations of the two programs and different input values for
water, temperature, and pressure profiles.
%
%A variation of typically 10\% between several mm clear-sky radiative
%transfer models was found in a systematic study \citep{melsheimer2005}
%with free input of spectroscopic data and continuum model.
%
Intercomparison of clear-sky radiative transfer models
\citep{melsheimer2005}, when the user had the free choice of selecting
spectroscopic data and continuum model for their calculation, show
difference of 10\% at mm wavelength.

% mainly due to the values of the spectroscopic parameters
%and absorption continuum description in each model.}

One advantage of the Mauna Kea site is that the water
line at 183 GHz \citep{wiedner2001}, and the continuum opacity at 225
GHz are continuosly monitored with dedicated radiometers. This allows
a precise calibration of data and to establish a helpful statistics on
the diurnal and yearly variation of the optical depth at 225 GHz
($\tau_{225}$). It was found that $\tau_{225}$ and the precipitable
water are related by $\tau_{225} = 0.01+0.04 \,{\rm pwv} $ and that
the opacity at higher frequencies, i.e. 345 GHz, can be calculated by
$\tau_{345} = 0.05+2.5\,
\tau_{225}$. The opacity at any frequency up to 1.6 THz can be 
accurately calculated using the WVM measurements at 183 GHz
\citep{pardo2004}.

\subsubsection{ALMA (5100m)} 

As stated on the website of the APEX radiotelescope
(http://www.apex-telescope.org/sites/chajnantor/atmosphere), the
median pwv is 1mm but can occasionally fall below 0.3mm.  Conditions
of pwv$<$0.5mm can be expected up to 25\% of the time between April
and December (the interval between December and April is used for
technical work and commissioning of new instruments and could 
potentially be considered for long-term atmospheric monitoring
projects).  We thus run models for pwvs of 0.3, 0.6, and 1mm and
display the results in Fig.~\ref{alma}. The absoute difference of
transmission compared to the Mauna Kea site is typically smaller than
10\% over all frequency ranges. What makes ALMA an exceptional site is
the high cumulative probablity of pwv values below 1mm
\citep{oterola2005}.

A first order comparison with the ATM model in the frequency range 10
to 1610 GHz (that is provided by a web-based calculator, see
http://www.apex-telescope.org/sites/chajnantor/atmosphere/transpwv/)
for the same pwv's values shows a typical deviation of 5-10\% between
the results. In general, the ATM model is more optimistic than
MOLIERE-5, but the general agreement is very good. Since the
comparison ATM/MOLIERE-5 for the CSO website was even better
(deviations of less than 5\%) it has to be clarified why the
difference for the ALMA site is larger. This is either due to the
temperature, pressure, and H$_2$O profiles and/or different versions
of the ATM model for Mauna Kea and ALMA site.

\subsubsection{DOME-C (3200m)} 

The transmission curves for the DOME-C site (Fig.~\ref{domec}) for
values of pwv=0.3, 0.6, and 1mm are typically 5--10\% lower than the
ones for ALMA. Below 1 THz, the difference is 10\% and above 1 THz the
difference becomes smaller with a transmission of up to 15\% for the
high-frequency windows for ALMA and 10\% for DOME-C.

Site-testing studies for DOME-C (see e.g. \citep{minier2008} for a
summary) show that the absolute values for pwv can drop very low. We
thus run models of pwvs of 0.1, 0.2, and 0.3 mm, shown in
Fig.~\ref{domec1}. For exceptional conditions, i.e. pwv around 0.1mm,
the transmission in some THz-windows of interest can go up to 30\%.
This is, however, also true for the ALMA site (5100m) and possible
future sites like that chosen for CCAT (5600m) or DOME-A (4100m).  In
fact, CCAT -- due to its higher altitude -- offers an even better
transmission at the same pwv than all other sites investigated
in this study (see webpage for more info). Recent, long-term
measurements at 200 $\mu$m [Tremblin et al. 2009, in prep.] show,
however, that Dome C is a better site than the best Chajnantor site
(i.e. mountains at $>$5500 m) in terms of pwv percentiles.

% There are only two future astronomical instruments that will observe
%in the THz range, this is HIFI onboard the Herschel satellite
%(http://herschel.esac.esa.int), but this instrument has a gap between
%1250 and 1410 GHz, and the NASA/DLR aircraft SOFIA
%(http://www.sofia.usra.edu) which is not yet commissioned. Earth-bound
%observatory sites for the astronomical THz range are thus lacking.

\begin{figure*}
  \begin{center}
   \includegraphics[width=9cm,angle=-90]{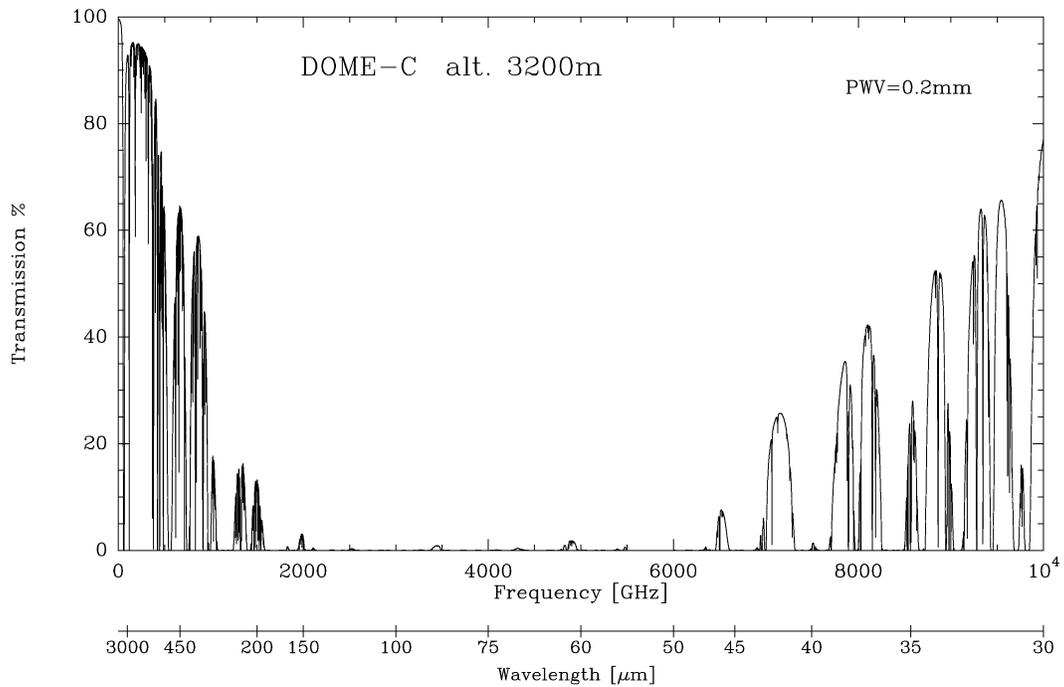}
  \end{center}
  \caption{Atmospheric transmission (in \%) for 0.2\,mm precipitable water
    (pwv) for the frequency range 0--10 THz for the DOME-C site
    (Concordia station/Antarctica). The frequency resolution is 200
    MHz. All H$_2$O and minor species lines up to 10 THz are included.}
  \label{domec2}
\end{figure*}

To explore in more detail the quality of the DOME-C site in the mid-IR
wavelength range, we run a model at pwv=0.2mm for frequencies up to 
10 THz ($\sim$3mm to 30 $\mu$m). Figure~\ref{domec2} demonstrates
that several windows open between 30 and 50 $\mu$m.
%Astronomical continuum observations in this wavelength range 
%of any astrophysical object, e.g. star forming regions, will allow to produce 
%a spectral energy distribution (SED), and thus 
%obtain the temperature under the assumption of blackbody emission.
%{\bf Nicola: Too detailed ? I can propose a shorter version:} 
Astronomical continuum observations in this wavelength range trace,
for example, warm dust in star forming regions. Combined with
continuum data in the FIR range (in the 200, 350, 450, and 870 $\mu$m
windows), this will allow to derive spectral energy distributions and
thus the temperature of astrophysical objects. Several bolometers are
already available for these wavelengths ranges: SHARC-II
\citep{dowell2001} at 350$\mu$m installed at the CSO, LABOCA
(http://www.mpifr-bonn.mpg.de/div/bolometer) at 870 $\mu$m at APEX,
and p-ARTEMIS \citep{talvard2006} at 450 $\mu$m \citep{andre2008} at
APEX (200 and 350 $\mu$m are planned for a future ARTEMIS
instrument). \\ \\

\section {Summary}

%Part 1 RT
We compiled basic properties and technical characteristics (optics,
front- and back-ends, observing modes, calibration) of radiotelescopes
with regard to their use for atmospheric line observations. Extensive
tables inform about telescope's location, instrumentation and
technical details as well as a web site for further information 
(http://transmissioncurves.free.fr).
We conclude that many radiotelescopes offer a good potential to be
used for atmospheric line observations and encourage the aeronomy
community to consider this potential.

%Part2 Moliere
The capacity of the MOLIERE-5 radiative transfer model was
demonstrated to produce transmission spectra for four astronomical
observatory sites (KOSMA, Mauna Kea, ALMA, DOME-C) for frequencies
between 0 and 2 THz (up to 10 THz for DOME-C). These transmission
curves are very useful to deduce the quality of a site for
astronomical or atmospheric observations in the millimeter to FIR
range. Results of model runs for other sites in form of graphical
displays and ascii-data are given on a dedicated website. \\ \\

%MOLIERE-5
%is continuously improved, the next development steps are (i) including
%all emission lines -- not only H$_2$O -- between 3 and 10 THz, (ii)
%extending the exploitable model range up to the IR range (30 to 4
%$\mu$m).... {\bf Nicola: What else, what are your plans ? } \\ 

{\bf Acknowlegement} 
We thank F. Boes for providing us with figures, and C. Kramer and
V. Minier for providing information concerning high-frequency spectral
lines and DOME-C site testing. We also thank D. Despois for 
useful discussions on HCN and radiotelescopes. 

%
%
%
%-------------------------------------------------------------------------

%-------------------------------------------------------------------------
% Table 1 
\begin{table*}[ht]
\caption{Sites of (sub)millimeter-wave radiotelescopes.} 
\renewcommand{\arraystretch}{1.3}
\begin{flushleft} 
\begin{tabular}{lclcclc} 
%\tableline 
Site        & $\varnothing$ & Location    & Geographical           & Altitude & Website  &  \\ 
            &  [m]     &                  & Longitude/Latitude     & [m]      & http://  &  \\ 
\hline 
CfA$^a$     & 1.2 & Cambridge/USA         & $\sim$74$^\circ$00$'$W, 40$^\circ$42$'$N            & 6    & cfa-www.harvard.edu/cfa/mmw & \\
%AST/RO$^b$  & 1.7 & South Pole/Antarctica & 180$^\circ$W,90$^\circ$S                            & 2835 & cfa-www.harvard.edu/$\sim$adair/AST$\_$RO &\\
KOSMA$^b$   & 3   & Gornergrat/Switzerland   &   7$^\circ$47$'$42$''$E, 45$^\circ$59$'$2$''$N      & 3200 & www.ph1.uni-koeln.de/kosma &\\
NANTEN2     & 4   & Pampa la Bola/Chile   & $\sim$67$^\circ$45$'$E, 23$^\circ$00$'$S            & 4865 & www.ph1.uni-koeln.de/nanten2 &\\
SRAO        & 6   & Seoul/Korea           & $\sim$37$^\circ$27$'$15$''$E, 126$^\circ$57$'$19$''$S  & 181 & srao.snu.ac.kr &\\
SMT$^c$     & 10  & Mt. Graham/USA        & 109$^\circ$53$'$28.5$''$W, 32$^\circ$42$'$5.8$''$N  & 3186 & aro.as.arizona.edu & \\ 
ASTE$^d$    & 10  & Pampa La Bola/Chile   &   67$^\circ$42$'$11$''$W,22$^\circ$58$'$18$''$S       & 4860 & www.nro.nao.ac.jp/~aste & \\ 
CSO$^e$     & 10  & Mauna Kea/Hawaii      & 155$^\circ$28$'$47$''$W, 19$^\circ$49$'$33$''$N     & 4092 & www.submm.caltech.edu/cso & \\ 
APEX$^f$    & 12  & Chajnantor/Chile      &  67$^\circ$42$'$11$''$W, 23$^\circ$00$'$20.8$''$S   & 5105 & www.apex-telescope.org & \\
KP12m$^g$   & 12  & Kitt Peak/USA         & 111$^\circ$36$'$53.475$''$W, 31$^\circ$57$'$12$''$N & 1914 & aro.as.arizona.edu & \\
Qinghai     & 13.7& Delingha/China        & 97$^\circ$33$'$36$''$E, 37$^\circ$22$'$24$''$N      & 3200 & www.pmodlh.ac.cn/13.7m\_telescope & \\
Mets\"ahovi & 14  & Kylm\"al\"a/Finland   & 24$^\circ$23$'$35$''$E, 60$^\circ$13$'$4$''$N       &  60   & kurp-www.hut.fi & \\ 
JCMT$^h$    & 15  & Mauna Kea/Hawaii      & 155$^\circ$28$'$47$''$W, 19$^\circ$49$'$33$''$N     & 4092 & www.jach.hawaii.edu/JCMT & \\ 
OSO$^i$     & 20  & Onsala/Sweden         &  11$^\circ$55$'$34.9$''$E, 57$^\circ$23$'$45$''$N   & 23   & www.chalmers.se/rss/oso-en & \\ 
Mopra       & 22  & Coonabarabran/Australia & 31$^\circ$16$'$04$''$E, 149$^\circ$05$'$58$''$S & 866 & www.narrabri.atnf.csiro.au & \\ 
IRAM$^j$    & 30  & Sierra Nevada/Spain   &   3$^\circ$23$'$33.7$''$W, 37$^\circ$3$'$58.3$''$N  & 2920 & iram.fr & \\
NRO$^k$     & 45  & Nobeyama/Japan        & 142$^\circ$43$'$2$''$, 35$^\circ$56$'$29.5$''$N     & 1350 & www.nro.nao.ac.jp  & \\ 
\hline
%Southern Mini & 1.2  & Cerro Tololo/Chile &  &  & cfa-www.harvard.edu/cfa/mmw/tech.html & \\
%FCRAO$^g$  & 14  &  New Salem/USA            & & & www.astro.umass.edu/\%7Efcrao/ & \\ 
%RT-22 LPI Radio Telescope 
%LMT   & 50 & Sierra Negra/Mexico & xx$^\circ$xx$'$xx$''$W,19$^\circ$xx$'$xx$''$ & 4600  & & \\
%  
\label{table1} 
\end{tabular} 
\end{flushleft} 
$^a$Center for Astrophysics, 
%$^b$Antarctic Submillimeter Telescope and Remote Observatory, 
$^b$K\"olner Observatorium f\"ur Sub-Mm Astronomie, 
$^c$SubMillimeter Telescope, 
$^d$Atacama Submillimeter Telescope Experiment,  
$^e$Caltech Submillimeter Observatory, 
$^f$Atacama Pathfinder EXperiment,  
$^g$Kitt Peak 12 meter Telescope,  
$^h$James Clark Maxwell Telescope, 
$^i$Onsala Space Observatory, 
$^j$Institut Radio Astronomie Millimetrique, 
$^k$Nobeyama Radio Observatory.
\end{table*} 
%-------------------------------------------------------------------------
\newpage

%-------------------------------------------------------------------------
% Bibliography using *.bbl file created by bibtex:
%
%\begin{scriptsize}
\bibliography{nschneider_2009,jurban_2009}

\begin{thebibliography}{71}
\expandafter\ifx\csname natexlab\endcsname\relax\def\natexlab#1{#1}\fi
\expandafter\ifx\csname url\endcsname\relax
  \def\url#1{\texttt{#1}}\fi
\expandafter\ifx\csname urlprefix\endcsname\relax\def\urlprefix{URL }\fi

\bibitem[{Allan(1972)}]{allan1972}
Allan, D., June 1972. {Statistics of atomic frequency standards}, precision
  measurement and calibration: selected nbs papers on frequency and time.
  Edition. Vol.~5 of NBS Special Publication 300. US Goverment Printing Office
  - Washington, DC, pp. 466--, edited by B.E. Blair and A.H. Morgan.

\bibitem[{Andr\'e et~al.(2008)Andr\'e, Minier, Gallais, and et~al.}]{andre2008}
Andr\'e, P., Minier, V., Gallais, P., et~al., 2008. {First 450 $\mu$m dust
  continuum mapping of the massive star-forming region NGC3576 with the
  P-ArT\'eMiS bolometer array}. Astronomy \& Astrophysics 490, L27.

\bibitem[{Barret and Chung(1962)}]{barret1962}
Barret, A., Chung, V., 1962. {A method for the determination of high altitude
  water vapour abundance from ground-based mm-observations}. J. Geophys. Res.
  67, 4295--4266.

\bibitem[{Bevilaqua et~al.(1983)Bevilaqua, Olivero, and
  Schwartz}]{bevilaqua1983}
Bevilaqua, R., Olivero, J., Schwartz, P., 1983. {An observational study of
  water vapour using ground-based microwave techniques}. J. Geophys. Res. 88,
  8523--.

\bibitem[{Boes(1994)}]{boes1994}
Boes, F., 1994. {Kohlenstoffmonoxid in der Mesosph\"are}. Ph.D. thesis,
  University of Cologne.

\bibitem[{Boissoles et~al.(2003)Boissoles, Boulet, Tipping, Brown, and
  Ma}]{boissoles2003}
Boissoles, J., Boulet, C., Tipping, R., Brown, A., Ma, Q., 2003. {Theoretical
  calculation of the translation-rotation collision-induced absorption in
  N$_2$-N$_2$, O$_2$-O$_2$ and N$_2$-O$_2$ pairs}. Jour.of Quant.Spec. \&
  Radiative Transfer 1-4, 505--.

\bibitem[{Borysow and Frommhold(1986)}]{borysow1986}
Borysow, A., Frommhold, L., December 15 1986. {Collision-induced
  rototranslational absorption spectra of N$_2$-N$_2$ pairs for temperatures
  from 50 to 300\,K}. Astrophysical Journal 311, 1043--1057.

\bibitem[{Burrows et~al.(2007)Burrows, Martin, and Roberts}]{burrows2007}
Burrows, S., Martin, C., Roberts, E., 2007. {High-latitude remote sensing of
  mesospheric wind speeds and carbon monoxide}. J. Geophys. Res. 112, 134--.

\bibitem[{{Calisesi} et~al.(2001){Calisesi}, {Wernli}, and
  {K{\"a}mpfer}}]{calisesi2001}
{Calisesi}, Y., {Wernli}, H., {K{\"a}mpfer}, N., 2001. {Midstratospheric ozone
  variability over Bern related to planetary wave activity during the winters
  1994-1995 to 1998-1999}. J. Geophys. Res. 106, 7903--7916.

\bibitem[{Caton et~al.(1968)Caton, Manella, and Kalgahan}]{caton1968}
Caton, W., Manella, G., Kalgahan, P., 1968. {Radio measurements of the
  atmospheric ozone transition}. Astrophysical Journal 151, L153--L156.

\bibitem[{Clough et~al.(1989)Clough, Kneizys, and Davies}]{ckd1989}
Clough, S., Kneizys, F., Davies, R., 1989. Line shape and the water vapor
  continuum. Atmos. Res. 23, 229--241.

\bibitem[{Clough et~al.(2005)Clough, Shepard, and et~al.}]{clough2005}
Clough, S., Shepard, M., et~al., E.~M., 2005. {Atmospheric radiative transfer
  modeling}. Journal of Quant. Spec. and Rad. Transfer 91, 233.

\bibitem[{Connor et~al.(1987)Connor, Barrett, and Parrish}]{connor1987}
Connor, B., Barrett, J., Parrish, A., 1987. {Ozone over McMurdo Station}. J.
  Geophys. Res. 92, 13221--.

\bibitem[{Crewell et~al.(1995)Crewell, Cheng, de~Zafra, and
  Trimble}]{crewell1996}
Crewell, S., Cheng, D., de~Zafra, R., Trimble, C., 1995. Millimeter wave
  spectroscopic measurements over the south pole, {I}: {A} study of
  stratospheric dynamics using {N$_2$O} observations. J. Geophys. Res. 100,
  20839--20844.

\bibitem[{{de~Zafra} and Smyshlyaev(2001)}]{dezafra2001}
{de~Zafra}, R., Smyshlyaev, S., 16~October 2001. {On the formation of HNO$_3$
  in the Antarctic mid to upper stratosphere in winter}. J. Geophys. Res.
  106~(D19), 23115--23125.

\bibitem[{{de Zafra} and {Muscari}(2004)}]{dezafra2004}
{de Zafra}, R.~L., {Muscari}, G., March 2004. {CO as an important high-altitude
  tracer of dynamics in the polar stratosphere and mesosphere}. J. Geophys.
  Res. 109~(D18), 6105--+.

\bibitem[{Despois et~al.(2000)Despois, Ricaud, Lauti\'e, Schneider, Jaqc,
  Biver, Lis, Chamberlin, Phillips, Miller, and Jenniskens}]{despois2000}
Despois, D., Ricaud, P., Lauti\'e, N., Schneider, N., Jaqc, T., Biver, N., Lis,
  D., Chamberlin, R., Phillips, T., Miller, M., Jenniskens, P., 2000. {Search
  for extraterrestrial origin of atmospheric trace molecules - Radio sub-mm
  observations during the Leonids}. Earth, Moon, and Planets 82-83, 129--140,
  special issue: Leonid storm.

\bibitem[{Dowell et~al.(2001)Dowell, Collins, Gardner, and et~al.}]{dowell2001}
Dowell, C., Collins, W., Gardner, M., et~al., 2001. {SHARC II, a Second
  Generation 350 Micron Camera for the CSO}. Bulletin of the American
  Astronomical Society 33, 792.

\bibitem[{Encrenaz et~al.(2004)Encrenaz, Lellouch, Atreya, and
  Wong}]{encrenaz2004}
Encrenaz, T., Lellouch, E., Atreya, S., Wong, A., 2004. {Detectability of minor
  constituents in the martian atmosphere by IR and submm spectroscopy}.
  Planetary and Space Science 52, 1023--.

\bibitem[{Eriksson(1999)}]{eriksson1999}
Eriksson, P., 1999. Microwave radiometric observations of the middle
  atmosphere: {S}imulations and inversions. Ph.D. thesis, Chalmers University
  of Technology, G{\"o}teburg, Sweden, technical report No. 355, {School of
  Electrical and Computer Engineering, ISBN 91-7197-757-0}.

\bibitem[{Forkman et~al.(2005)Forkman, Eriksson, Murtagh, and
  Espy}]{forkman2005}
Forkman, P., Eriksson, P., Murtagh, D., Espy, P., 2005. {Observing the vertical
  branch of the mesospheric circulation at latitude 60$^\circ$N using
  ground-based measurements of CO and H$_2$O}. J. Geophys. Res. 110~(D05107),
  doi:10.1029/2004JD004916.

\bibitem[{Frisk et~al.(2003)Frisk, Hagstr\"om, Ala-Laurinaho, and
  et~al.}]{frisk2003}
Frisk, U., Hagstr\"om, M., Ala-Laurinaho, J., et~al., May 2003. {The Odin
  satellite: I. Radiometer design and test}. Astronomy \& Astrophysics 402~(3),
  L27--34.

\bibitem[{Groom(1965)}]{groom1965}
Groom, D., 1965. {Stratospheric thermal emission and absorption neat the
  1.35\,cm line of water vapour}. J. Atmos. Terr. Phys. 27, 217--233.

\bibitem[{Kasai et~al.(2006)Kasai, Takahashi, Urban, Hoshino, Takahashi,
  Inatani, Shiotani, and Masuko}]{kasai2006}
Kasai, Y., Takahashi, C., Urban, J., Hoshino, S., Takahashi, K., Inatani, J.,
  Shiotani, M., Masuko, H., March 2006. {Stratospheric Ozone Isotope Enrichment
  Studied by Sub-Millimeter Wave Heterodyne Radiometry: The Observation
  Capabilities of SMILES}. IEEE Transactions on Geoscience and Remote Sensing
  44~(3), 676--693.

\bibitem[{Klein et~al.(2002)Klein, Wohltmann, Lindtner, and
  K\"unzi}]{klein2002}
Klein, U., Wohltmann, I., Lindtner, K., K\"unzi, K., 2002. {Ozone depletion and
  chlorine activation in the Arctic winter 1999/2000 observed in Ny-Alesund}.
  J. Geophys. Res. 107~(D20), 8288--, doi:10.1029/2001JD000543.

\bibitem[{Kramer and Stutzki(1990)}]{kramer1990}
Kramer, C., Stutzki, J., 1990. {Atmospheric Transparency at Gornergrat, Univ.
  of Cologne}. Technical Mem. 5.

\bibitem[{Kraus(1986)}]{kraus1986}
Kraus, J., 1986. {Radio Astronomy}. Cygnus-Quasar Books. Powell.

\bibitem[{{Lauti\'e}(2003)}]{lautie2003}
{Lauti\'e}, N., 2003. {Traitement des mesures satellitaires
  sub-millim\'etriques effectu\'ees par {Odin/SMR}; \'etude non-lin\'eaire de
  la vapeur d'eau. \'Etude stratosph\'erique de {HCN} au moyen de mesures
  micro-ondes.} Ph.D. thesis, Universit\'e Paris VI, France.

\bibitem[{Lawrence et~al.(2004)Lawrence, Ashley, Tokovinin, and
  Travouillon}]{lawrence2004}
Lawrence, J., Ashley, M., Tokovinin, A., Travouillon, T., 2004. {Exceptional
  astronomical seeing conditions above DOME-C in Antarctica}. Nature 431.

\bibitem[{Liebe et~al.(1993)Liebe, Hufford, and Cotton}]{liebe1993}
Liebe, H., Hufford, G., Cotton, M., 17-21 May 1993. Propagation modeling of
  moist air and suspended water/ice particles at frequencies below 1000\,{GHz},
  {AGARD} 52nd Specialists Meeting of the Electromagnetic Wave Propagation
  Panel, Palma De Mallorca, Spain.

\bibitem[{Lobsiger et~al.(1984)Lobsiger, K\"unzi, and D\"utsch}]{lobsinger1984}
Lobsiger, E., K\"unzi, K., D\"utsch, H., 1984. {Comparison of stratosperic
  ozone profiles retrieved from microwave radiometer and Dobson spectrometer
  data}. J. Atm. and Terr. Phys. 46~(9), 799.

\bibitem[{Lord(1992)}]{lord1992}
Lord, S., 1992. {A new software tool for computing earth's atmospheric
  transmission of near- and far-infrared radiation}. NASA Technical
  Memorandum~(103957).

\bibitem[{Marks(2002)}]{marks2002}
Marks, R., 2002. {Astronomical seeing from the summits of the Antarctic
  Plateau}. Astronomy \& Astrophysics 385, 328.

\bibitem[{Melsheimer et~al.(2005)Melsheimer, Verdes, Buehler, and
  et~al.}]{melsheimer2005}
Melsheimer, C., Verdes, C., Buehler, S., et~al., 2005. Intercomparison of
  general purpose clear-sky atmospheric radiative transfer models for the
  millimeter/sub-millimeter spectral range. Radio Science 40~(RS1007),
  doi:10.1029/2004RS003110.

\bibitem[{Minier et~al.(2008)Minier, Olmi, Lagage, and et~al.}]{minier2008}
Minier, V., Olmi, L., Lagage, P.-O., et~al., 2008. {Submm/FIR astronomy in
  Antarctica: Potential for a large telescope facility}. In: Science, E. (Ed.),
  {Proc. of the 2nd ARENA Conference, Potsdam, Vol 512}.

\bibitem[{Muscari et~al.(2002)Muscari, Santee, and {de~Zafra}}]{muscari2002}
Muscari, G., Santee, M., {de~Zafra}, R., 2002. {Intercomparison of
  stratospheric HNO$_3$ measurements over Antarctica: Ground-based
  millimeter-wave versus UARS/MLS Version 5 retrievals}. J. Geophys. Res.
  107~(D24), 4809--.

\bibitem[{Nedoluha et~al.(2003)Nedoluha, Bevilacqua, Gomez, Hicks,
  {Russell~III}, and Connor}]{nedoluha2003}
Nedoluha, G., Bevilacqua, R., Gomez, R., Hicks, B., {Russell~III}, J., Connor,
  B., 10 July 2003. {An evaluation of trends in middle atmospheric water vapor
  as measured by HALOE, WVMS, and POAM}. J. Geophys. Res. 108~(D13), 4391,
  doi:10.1029/2002JD003332.

\bibitem[{Oterola et~al.(2005)Oterola, Holdaway, Nyman, Radford, and
  Butler}]{oterola2005}
Oterola, A., Holdaway, M., Nyman, L.-E., Radford, S., Butler, B., 2005.
  {Atmospheric Transparance at Chanjanantor: 1973-2003}. ALMA mem. Series 512.

\bibitem[{Paine(2004)}]{paine2004}
Paine, S., 2004. {The AM atmospheric model}. SMA Technical Memorandum~(152).

\bibitem[{Pardo et~al.(2001)Pardo, Cernicharo, and Serabyn}]{pardo2001}
Pardo, J., Cernicharo, J., Serabyn, E., 2001. Atmospheric transmission at
  microwaves (atm): an improved model for mm/submm applications. IEEE
  Transactions on Antennas and Propagation 49~(12), 1683--.

\bibitem[{Pardo et~al.(1995)Pardo, Pagani, G\'erin, and Prigent}]{pardo1995}
Pardo, J., Pagani, L., G\'erin, M., Prigent, C., 1995. {Evidence of Zeeman
  splitting in the 21-01 rotational transition of atmosp. $^{16}$O$^{18}$O}. J.
  Quant. Spectrosc. Radiat. Transfer 54~(6), 931--.

\bibitem[{Pardo et~al.(2005)Pardo, Serabyn, Wiedner, and
  Cernicharo}]{pardo2005}
Pardo, J., Serabyn, E., Wiedner, M., Cernicharo, J., 2005. {Measured telluric
  continuum-like opacity beyond 1 THz}. Jour. of Quant. Spec.\& Rad. Transfer
  96, 537--.

\bibitem[{Pardo et~al.(2004)Pardo, Wiedner, Serabyn, and et~al.}]{pardo2004}
Pardo, J., Wiedner, M., Serabyn, E., et~al., 2004. Side-by-side comparison of
  fts and wvr as tools for the calibration of mm/submm ground-based
  observatories. Astrophysical Journal Supplement 153, 363.

\bibitem[{Parrish et~al.(1981)Parrish, {de~Zafra}, and Solomon}]{parrish1981}
Parrish, A., {de~Zafra}, R., Solomon, P., 1981. {ClO in the stratospheric ozone
  layer: Ground-based detection and measurement}. Science 211, 1158--.

\bibitem[{Parrish et~al.(1988)Parrish, {de~Zafra}, Solomon, and
  Barrett}]{parrish1988}
Parrish, A., {de~Zafra}, R., Solomon, P., Barrett, J., 1988. {A ground based
  technique for millimeter wave measurements of stratospheric trace
  constituents}. Radio Science 23~(2), 106--118.

\bibitem[{Penzias and Burrus(1973)}]{penzias1973}
Penzias, A., Burrus, C., 1973. {Millimeter-Wavelength Radio-Astronomy
  Techniques}. Annual Review of Astronomy and Astrophysics 11, 51--.

\bibitem[{Pickett et~al.(1998)Pickett, Poynter, Cohen, Delitsky, Pearson, and
  M{\"u}ller}]{pickett1998}
Pickett, H., Poynter, R., Cohen, E., Delitsky, M., Pearson, J., M{\"u}ller, H.,
  1998. Submillimeter, millimeter, and microwave spectral line catalog. J.
  Quant. Spectrosc. Radiat. Transfer 60~(5), 883--890.

\bibitem[{Prigent et~al.(2006)Prigent, Pardo, and Rossow}]{prigent2006}
Prigent, C., Pardo, J., Rossow, W., 2006. {Comparisons of the mm and submm
  bands for atmospheric temperature and water vapor soundings from clear and
  cloudy skies}. Journal of Applied Meteorology and climatology 45, 1622--.

\bibitem[{{Raffalski} et~al.(2005){Raffalski}, {Hochschild}, {Kopp}, and
  {Urban}}]{raffalski2005}
{Raffalski}, U., {Hochschild}, G., {Kopp}, G., {Urban}, J., 2005. {Evolution of
  stratospheric ozone during winter 2002/2003 as observed by a ground-based
  millimetre wave radiometer at Kiruna, Sweden}. Atmospheric Chemistry \&
  Physics 5, 1--9.

\bibitem[{Ricaud et~al.(2004)Ricaud, Baron, and {de~La~ No\"e}}]{ricaud2004}
Ricaud, P., Baron, P., {de~La~ No\"e}, J., 2004. {Quality assessment of
  ground-based microwave measurements of ClO, O$_3$, NO$_2$ from the NDSC
  radiometer at the Plateau de Bure}. Annales Geophysicae 22, 1903--1915.

\bibitem[{Ricaud et~al.(1991)Ricaud, Brillet, {de~La~ No\"e}, and
  Parisot}]{ricaud1991}
Ricaud, P., Brillet, J., {de~La~ No\"e}, J., Parisot, J., 1991. {Diurnal and
  seasonal variations of stratomesopheric ozone}. J. Geophys. Res. 96, 18617--.

\bibitem[{Rosenkranz(1993)}]{rosenkranz1993}
Rosenkranz, P., 1993. Absorption of Microwaves by Atmospheric Gases,
  {Atmospheric Remote Sensing by Microwave Radiometry} Edition. Wiley Series in
  Remote Sensing. M.A. Janssen, New York, Ch.~2, pp. 37--82, {ISBN
  0-471-62891-3}.

\bibitem[{Rothman et~al.(1998)Rothman, Rinsland, Goldman, and
  et~al.}]{rothman1998}
Rothman, L., Rinsland, C., Goldman, A., et~al., November 1998. {The HITRAN
  Molecular Spectroscopic Database and HAWKS (HITRAN Atmospheric Workstation):
  1996 Edition}. J. Quant. Spectrosc. Radiat. Transfer 60, 665--710.

\bibitem[{Sandor and Clancy(1998)}]{sandor1998}
Sandor, B., Clancy, R., 1998. {Mesospheric HO$_2$ chemistry}. J. Geophys. Res.
  103.

\bibitem[{Schneider et~al.(2003)Schneider, Lezeaux, {de~La~No{\"e}}, Urban, and
  Ricaud}]{schneider2003}
Schneider, N., Lezeaux, O., {de~La~No{\"e}}, J., Urban, J., Ricaud, P., 2003.
  Validation of ground-based strato-mesospheric ozone observations. J. Geophys.
  Res. 108~(D17), 4540+.

\bibitem[{{Seele} and {Hartogh}(1999)}]{seele1999}
{Seele}, C., {Hartogh}, P., June 1999. {Water vapor of the polar middle
  atmosphere: Annual variation and summer mesosphere conditions as observed by
  ground-based microwave spectroscopy}. Geophys. Res. Lett. 26, 1517--1520.

\bibitem[{{Shindell} and {de Zafra}(1996)}]{shindell1996}
{Shindell}, D.~T., {de Zafra}, R.~L., 1996. {Chlorine monoxide in the Antarctic
  spring vortex 2. A comparison of measured and modeled diurnal cycling over
  McMurdo Station, 1993}. J. Geophys. Res. 101, 1475--1488.

\bibitem[{Sinnhuber et~al.(1998)Sinnhuber, Langer, Klein, Raffalski, K\"unzi,
  and Schrems}]{sinnhuber1998}
Sinnhuber, B.-M., Langer, J., Klein, U., Raffalski, U., K\"unzi, K., Schrems,
  O., 1998. Ground based millimeter-wave observations of {A}rctic ozone
  depletion during winter and spring 1996/97. Geophys. Res. Lett. 25~(17),
  3327--3330.

\bibitem[{Solomon et~al.(1984)Solomon, {de~Zafra}, and Parrish}]{solomon1984}
Solomon, P., {de~Zafra}, R., Parrish, A., 1984. {Diurnal variations of
  stratospheric ClO}. Science 224, 1210--.

\bibitem[{{Steinbrecht} et~al.(2006){Steinbrecht}, {Claude}, {Sch{\"o}nenborn},
  {McDermid}, {Leblanc}, {Godin}, {Song}, {Swart}, {Meijer}, {Bodeker},
  {Connor}, {K{\"a}mpfer}, {Hocke}, {Calisesi}, {Schneider}, {de~La~No\"e},
  {Parrish}, {Boyd}, {Br{\"u}hl}, {Steil}, {Giorgetta}, {Manzini}, {Thomason},
  {Zawodny}, {McCormick}, {Russell}, {Bhartia}, {Stolarski}, and
  {Hollandsworth-Frith}}]{steinbrecht2006}
{Steinbrecht}, W., {Claude}, H., {Sch{\"o}nenborn}, F., {McDermid}, I.~S.,
  {Leblanc}, T., {Godin}, S., {Song}, T., {Swart}, D.~P.~J., {Meijer}, Y.~J.,
  {Bodeker}, G.~E., {Connor}, B.~J., {K{\"a}mpfer}, N., {Hocke}, K.,
  {Calisesi}, Y., {Schneider}, N., {de~La~No\"e}, J., {Parrish}, A.~D., {Boyd},
  I.~S., {Br{\"u}hl}, C., {Steil}, B., {Giorgetta}, M.~A., {Manzini}, E.,
  {Thomason}, L.~W., {Zawodny}, J.~M., {McCormick}, M.~P., {Russell}, J.~M.,
  {Bhartia}, P.~K., {Stolarski}, R.~S., {Hollandsworth-Frith}, S.~M., 2006.
  {Long-term evolution of upper stratospheric ozone at selected stations of the
  Network for the Detection of Stratospheric Change (NDSC)}. J. Geophys. Res.
  111~(D10308), 10308--.

\bibitem[{Talvard et~al.(2006)Talvard, Andr\'e, Rodriguez, and
  et~al.}]{talvard2006}
Talvard, M., Andr\'e, P., Rodriguez, L., et~al., 2006. {ARTEMIS: filled
  bolometer arrays for next generation telescopes}. SPIE 6275.

\bibitem[{Urban et~al.(2004a)Urban, Baron, Lauti\'e, Dassas, Schneider, Ricaud,
  and de~La~No\"e}]{urban2004a}
Urban, J., Baron, P., Lauti\'e, N., Dassas, K., Schneider, N., Ricaud, P.,
  de~La~No\"e, J., 2004a. {MOLIERE (v5): A versatile forward- and inversion
  model for the millimeter and sub-millimeter wavelength range}. J. Quant.
  Spectrosc. Radiat. Transfer 83~(3-4), 529--554.

\bibitem[{Urban et~al.(2005a)Urban, Dassas, Ricaud, and Forget}]{urban2005a}
Urban, J., Dassas, K., Ricaud, P., Forget, F., April 2005a. Retrieval of
  vertical constituents and temperature profiles from passive sub-millimeter
  wave limb observations of the {M}artian atmosphere: a feasibility study.
  Applied Optics 44, 2438--2455.

\bibitem[{Urban et~al.(2000a)Urban, K{\"u}llmann, K{\"u}nzi, Wohlgemuth, Goede,
  Kleipool, Whyborn, Schwaab, and Chipperfield}]{urban2000a}
Urban, J., K{\"u}llmann, K., K{\"u}nzi, K., Wohlgemuth, J., Goede, A.,
  Kleipool, Q., Whyborn, N., Schwaab, G., Chipperfield, M., 2000a.
  Stratospheric {ClO} across the edge of the of the {A}rctic polar vortex:
  {M}easurements of the
  {A}ir\-borne-\-{S}ub\-milli\-meter-\-{SIS}-\-{R}adio\-meter compared to {3-D}
  model calculations. In: Zerefos, C. (Ed.), Chemistry and Radiation Changes in
  the Ozone layer. Vol. 557 of Nato Science Series {C}. {Kluwer Academic
  Publishers}, pp. 233--240, {ISBN 0-7923-6513-5}.

\bibitem[{Urban et~al.(2005b)Urban, Lauti\'e, {Le~Flochmo\"en}, Jim\'enez,
  Eriksson, Dupuy, {El~Amraoui}, Ekstr\"om, Frisk, Murtagh, de~La~No\"e,
  Olberg, and Ricaud}]{urban2005b}
Urban, J., Lauti\'e, N., {Le~Flochmo\"en}, E., Jim\'enez, C., Eriksson, P.,
  Dupuy, E., {El~Amraoui}, L., Ekstr\"om, M., Frisk, U., Murtagh, D.,
  de~La~No\"e, J., Olberg, M., Ricaud, P., July 2005b. Odin/smr limb
  observations of stratospheric trace gases: Level~2 processing of clo, n$_2$o,
  o$_3$, and hno$_3$. J. Geophys. Res. 110, D14307.

\bibitem[{Urban et~al.(2007a)Urban, Lauti\'e, Murtagh, Eriksson, Kasai,
  Lo{\ss}ow, Dupuy, {de\,La\,No\"e}, Frisk, Olberg, {Le~Flochmo\"en}, and
  Ricaud}]{urban2007a}
Urban, J., Lauti\'e, N., Murtagh, D., Eriksson, P., Kasai, Y., Lo{\ss}ow, S.,
  Dupuy, E., {de\,La\,No\"e}, J., Frisk, U., Olberg, M., {Le~Flochmo\"en}, E.,
  Ricaud, P., June 2007a. {Global observations of middle atmospheric water
  vapour by the Odin satellite: An overview.} Planetary and Space Science
  55~(9), 1093--1102, special issue 2nd General Assembly of Asia Oceania
  Geophysical Society (2005): Highlights in Planetary Science.

\bibitem[{Waters et~al.(2006)Waters, Froidevaux, Harwood, and
  et~al.}]{waters2006}
Waters, J., Froidevaux, L., Harwood, R., et~al., 2006. {The Earth observing
  system microwave limb sounder on the Aura satellite}. IEEE Trans. Geosci.
  Remote Sensing 44~(5).

\bibitem[{Waters et~al.(1976)Waters, Wilson, and Shimabukuro}]{waters1976}
Waters, J., Wilson, T., Shimabukuro, F., 1976. {Microwave measurements of
  mesospheric CO}. Science 192, 1174.

\bibitem[{Wieching(2007)}]{wieching2007}
Wieching, G., 2007. {CONDOR: a Heterodyne Receiver for Astronomical
  Observations at 1.5 THz}. PhD Thesis, University of Cologne.

\bibitem[{Wiedner et~al.(2001)Wiedner, Hills, Carlstrom, and
  et~al.}]{wiedner2001}
Wiedner, M., Hills, R., Carlstrom, J., et~al., 2001. {Interferometric Phase
  Correction using 183 GHz Water Vapor Monitors}. Astrophysical Journal 553,
  1039.

\bibitem[{Wiedner et~al.(2006)Wiedner, Wieching, and et~al.}]{wiedner2006}
Wiedner, M., Wieching, G., et~al., F.~B., 2006. {First observations with
  CONDOR: a 1.5 THz Heterodyne Receiver }. Astronomy \& Astrophysics 454, L33.

\end{thebibliography}
%\end{scriptsize}
%-------------------------------------------------------------------------
\newpage

%-------------------------------------------------------------------------
% Table 2
% Characteristics of mm-wave Radiotelescopes 
\begin{table*}[ht] 
\caption{Technical Specifications of Radiotelescopes.} 
\renewcommand{\arraystretch}{1.3}
\begin{flushleft} 
\begin{tabular}{lcccccccccc} 
Site & Freq. Range$^a$ & Type/Band$^b$ & T$_{rec}^c$  
& BW$_{rec}^d$ & IF$^e$ & Spec.$^f$ & BW$_{spec}^g$  & Spectral res. & Obs. Mode$^h$ & FWHM$^i$  \\ 
 & [GHz] &  & [K] & [GHz]  & [GHz] &  & [MHz] & [kHz] & & [$''$] \\ 
CfA     & 115      & SIS/SSB & 65       & 1   & 1.4 & FB  & 128           & 250,500  & PS,FS       & 480 \\
\hline
%AST/RO  & 230      & SIS/DSB & 100      & 1   & 1.5 & AOS & 64, 1100      & 44,1070  & PS,BS,FS    & 185 \\ 
%        & 460--492 & SIS/DSB & 220      & 1   & 1.5 &     &               &          & PS,LS,FS    & 109--103 \\ 
%        & 800--820 & SIS/DSB & 990      & 1   & 1.5 &     &               &          & PS,LS,FS    &  58 \\ 
%\hline
KOSMA   & 210--270 & SIS/DSB & 150      & 1   & 1.4 & AOS & 60--1000      & 50--1100 & PS,LS,BS,FS & 130 \\ 
        & 330--365 & SIS/DSB & 100      & 1   & 1.4 &     &               &          & PS,LS,BS,FS &  80 \\
\hline 
NANTEN2 & 455--495 & SIS/DSB & 150      & 1   & 1.5 & AOS & 4$\times$1000 &          & PS,LS,BS,FS &  40 \\ 
        & 795--881 & SIS/DSB & 500--700 & 1   & 1.5 &     &               &          & PS,LS,BS,FS &  22 \\ 
%        & 115      & SIS     &          &     &     &     &               &          &             & 180 \\ 
%        & 230      & SIS     &          &     &     &     &               &          &             &  90 \\ 
%        & 345      & SIS     &          &     &     &     &               &          &             &  60 \\ 
\hline
SRAO    & 110/230  & SIS/SSB &  60      &     &     & AC  & 50/100        & 12--50   & PS,BS       &  48 \\ 
\hline
SMT     & 130--300 & SIS/SSB &          & 2   &   5 & AOS & 1000          & 900      & PS,BS       &  58--25\\ 
        & 320--375 & SIS/DSB & 125      & 1   & 1.5 & FB  & 250--2048     & 385,1000 & PS,BS       &  22 \\ 
        & 425--500 & SIS/DSB & 110--150 & 0.8 & 4--6 & CTS & 215          & 40       & PS,BS       & 16.5 \\ 
        & 652--720 & SIS/DSB &  70--95  &     &     &     &               &          & PS,BS       &  11 \\ 
\hline
ASTE    & 324--366 & SIS/DSB & 125      &  4  & 4--8 & FB  & 128, 512     & 125,500  &             &  22 \\ 
        &          &         &          &     &      & FFTS & 4000        & 2000     &             & \\       
\hline
CSO     & 195--280 & SIS/DSB &  50      & 1   & 1.2--1.8 & AOS & 50-1500  & 50--700  & PS,BS,FS    & 33 \\
        & 280--420 & SIS/DSB &  50      & 1   & 1.2--1.8 & FFTS&500-1000  &          & PS,BS,FS    & 24 \\
        & 440--520 & SIS/DSB & 150      & 1   & 1.2--1.8 &     &          &          & PS,BS,FS    & 16.5 \\
        & 600--720 & SIS/DSB & 150      & 1   & 1.2--1.8 &     &          &          & PS,BS,FS    & 14.5 \\
        & 780--950 & SIS/DSB & 250      & 1   & 1.2--1.8 &     &          &          & PS,BS,FS    & 11.5 \\
\hline
APEX    & 211--275 & SIS/SSB &  80      &     & 4--8     & FFTS& 1000     & $>$65    & PS,FS       & 30--25 \\
        & 275--370 & SIS/SSB &  135     &     & 4--8     &     &          &          & PS,FS       & 23--17 \\
%        & 279--381 & SIS/DSB &  135     &     & 4--8     &     &          &          & PS,FS       &  18  \\
%        & 375--500 & SIS/SSB & 200      &     & 4--8     &     &          &          &             &  17--13 \\
        & 1250--1384 & SIS/DSB & 200    &     & 2--4     &     &          &          & PS, FS      &   5 \\
%        & 780--887 & SIS/DSB & 400      &     & 4--8     &     &          &          &             &   7 \\
%        & 602--720 & SIS/DSB &          &     &          &     &          &          &             & 9--7 \\	 
%        & 790--950 & SIS/DSB &          &     &          &     &          &          &             & 7--6 \\	 
\hline 
KP12m   & 68--90   & SIS/SSB & 100      & 1   & 1.5      & FB  & 3.8--512 & 30--2000 & PS,BS,FS    & 90--70 \\
        & 90--116  & SIS/SSB &  80      & 1   & 1.5      & AC  & 75--600  & 6--3125  & PS,BS,FS    & 70--55 \\
        & 133--180 & SIS/SSB & 125      & 1   & 1.5      &     &          &          & PS,BS,FS    & $\sim$45\\
\hline
Qinghai & 85--115  & SIS/SSB & 60       &     &          & AOS & 42--145  & 75--209  & PS          & 106$\times$70 \\
\hline
Mets\"ahovi & 80--115 & Schottky/DSB & 150 &  & 1--1.6   & AOS & 50--1500 & 50--700  & PS,BS,FS    & $\sim$45\\
        & 84--115  & SIS     & 100      &     & 3.7--4.2 &     &          &          & PS,BS,FS    & $\sim$45\\
        & 129--175 & SIS     & 150      &     & 3.7--4.1 &     &          &          & PS,BS,FS    & $\sim$37\\
        & 147      & SIS     & 145      &     & 3.7--4.2 &     &          &          & PS,BS,FS    & $\sim$37 \\
        & 280--420 & SIS/DSB &  50      & 1   & 1.2--1.8 &     &          &          & PS,BS,FS    & 24 \\
\hline
JCMT    & 211--279 & SIS/DSB &  70      & 1.8 & 4        & AC & 200--1800 & 30.5--977 & PS,BS,FS  & 20\\ 
        & 325--375 & SIS/DSB & 150      & 1.8 & 4        &     &          &         & PS,BS,FS     & 14 \\ 
        & 430--510 & SIS/DSB & 330      & 1.6 & 4        &     &          &         & PS,BS,FS     & 11 \\ 
        & 626--710 & SIS/DSB & 33       & 1.6 & 4        &     &          &         & PS,BS,FS     & 8 \\ 
%        & 870      & SIS/DSB &          &     &          &     &          &         & PS,BS,FS     & 6 \\ 
\hline
OSO     & 18--50   & Schottky& 30--50   &     &         & AC  & 0.05--1280 & 0.03--800 & PS,BS,FS & 210--75\\ 
        & 84--116  & SIS/SSB & 80--130  &     &         &     &            &  & PS,BS,FS & 44--33\\ 
\hline
MOPRA   & 30--50   & SIS/SSB & 30    &     &          & FB    & 8000        &         & PS,FS        & 115--69 \\    
     %  & 16--26   & SIS/SSB &       &     &          & FB  & 8000     &         & PS,FS        &  \\    
         & 76--117  & SIS/SSB & 100      &     &          &     &          &         & PS,FS        & 45--30 \\    
\hline
IRAM    & 80--115  & SIS/DSB & 60--80   & 0.5 & 1.5      & FB  & 10--1000 & 3--4000 & PS,BS,FS  & 29--22\\ 
        & 130--183 & SIS/DSB & 70--125  & 1   & 4        &     &          &         & PS,BS,FS     & 14--15 \\ 
        & 197--266 & SIS/DSB & 85--160  & 1   & 4        &     &          &         & PS,BS,FS     & 13--9 \\ 
        & 241--281 & SIS/DSB & 125--250 & 1   & 4        &     &          &         & PS,BS,FS     & 11--9 \\ 
        & 210--276 & SIS/DSB & 110--380 & 1   & 4        &     &          &         & PS,BS,FS     & 12--9 \\ 
\hline
NRO     &  72--115 & SIS/DSB & 250--900$^j$ & 0.6 &          & AOS & 40,250,480 & 1,37,250 & PS,FS  & 23--15 \\ 
        &  82--116 & SIS/DSB & 400--800$^j$ & 0.6 & 2--2.6   & AC  & 4--512   & 4--500  &           & 20--14 \\ 
\hline   
%LMT 50m   & 85--115  & DSB & SIS & 50--80  & 15 &  & AC & 1-2000 & 10--10$^5$ & PS,BS,FS & 10--5\\ 
%LMT sequioa array (4x4) receiver under construction 
\label{table2} 
\end{tabular} 
\end{flushleft} 
$^a$Tunable frequency range of the receiver. 
$^b$Type of receiver (SIS or Schottky)/Double-Sideband (DSB) or
Single-Sideband (SSB) operation. Some of the receivers are of
multi-beam type (not explicitly indicated). 
$^c$Receiver noise temperature (without atmosphere). 
$^d$Receiver bandwidth. 
$^e$Intermediate frequency. 
$^f$Spectrometer type AOS: acousto optical spectrometer,
    FB:filterbank, AC: autocorrelator, FFTS: Fast Fourier Transform Spectrometer.   
$^g$Spectrometer bandwidth.   
$^h$Observing Modes: PS=Position Switch, LS=Load Switch, BS=Beam Switch, FS=Frequency Switch. 
$^i$Beam Full Width Half Maximum for lower and upper frequency limits. 
%$^j$system temperature
%$^k$50,500 and 1500 MHz. 
%$^l$50,500, and 700 kHz channel spacing at the bandwidths given in $^k$. 
%$^m$125,250,500,920 and 1800 MHz. 
%$^n$95,189,378,756 and 1513 kHz at the bandwidths given in $^m$. 
%$^o$4x256,2x512or 1x1000 MHz, 2x1GHz, 2x12.8 or 1x25.6 MHz, 10--512 MHz. 
%$^p$3.3 kHz-1.25 MHz, 100 kHz, 1 and 4 MHz. 
%$^q$Beamsize for lower and upper frequency limits. 
\end{table*} 
%-------------------------------------------------------------------------
%\newpage

%-------------------------------------------------------------------------
% SOUNDINGS
%\begin{figure*}
%  \begin{center}
%  \includegraphics[width=10cm,angle=0]{soundings.eps}
%  \end{center}
%  \caption{Sketch of different observing geometries.}
%   \label{soundings}
%\end{figure*}

% FS

% LS
%\begin{figure*}
%  \begin{center}
%  \includegraphics[width=10cm,angle=0]{load-switch1.eps}
%  \end{center}
%  \caption{Example of a load swichting observation of $^{12}$CO 2$\to$1 at 
%  230 GHz, taken  with the KOSMA 3m telescope \citet{boes1994}.}
%  \label{ls}
% \end{figure*}

% ABS-SUN
%\begin{figure*}
%  \begin{center}
%  \includegraphics[width=10cm,angle=0]{sun-abs.eps}
%  \end{center}
%  \caption{Example of an absorption measurement against the sun of $^{12}$CO 2$\to$1 at 
%   230 GHz, taken with the KOSMA 3m telescope \citet{boes1994}.}
%   \label{abs-sun}
%\end{figure*}

% ABS-MOUNTAIN
%\begin{figure*}
%  \begin{center}
%  \includegraphics[width=10cm,angle=0]{mountain-abs.eps}
%  \end{center}
%  \caption{Example of a absorption measurement against the mountains of $^{12}$CO 2$\to$1 at 
%   230 GHz, taken with the KOSMA 3m telescope \citet{boes1994}.}
%   \label{abs-mountain}
%\end{figure*}

% DRY AIR 

% DRY AIR 
%\begin{figure*}
%  \begin{center}
%  \includegraphics[width=8cm,angle=-90]{trans_paper_wet.eps}
%  \end{center}
%  \caption{Transmission calculated for three different altitudes,
%   {\bf only taking into account} the wet component (line and continuum), 
%   based on the model of \citet{pardo2001}. A common pwv of 0.3mm is 
%   used. The water continuum alone is included for comparison. 
%   The temperature and pressure profiles are from the 45$^\circ$ N U.S. standard 
%   atmosphere, the spectral resolution is 200 MHz.}
%   \label{wet-air}
%\end{figure*}

% Figures with TRANSMISSION 
%-------------------------------------------------------------------------
%-------------------------------------------------------------------------
%-------------------------------------------------------------------------

%-------------------------------------------------------------------------
\end{document}